\documentclass[a4paper,11pt]{article}
\usepackage{jcappub} % for details on the use of the package, please
\usepackage[T1]{fontenc} % if needed

\usepackage{color}

\def\l{\left}
\def\r{\right}
\def\be{\begin{equation}}
\def\ee{\end{equation}}
\def\ba{\begin{eqnarray}}
\def\ea{\end{eqnarray}}
\def\bl#1\el{\begin{align}#1\end{align}}

\title{\boldmath Sudden braking and turning with a two-field potential bump: primordial black hole formation}

\author[a]{Chengjie Fu}

\author[b]{Chao Chen}

\affiliation[a]{Department of Physics, Anhui Normal University, Wuhu, Anhui 241000, P.R.China}

\affiliation[b]{Jockey Club Institute for Advanced Study, The Hong Kong University of Science and Technology, Hong Kong S.A.R., P.R.China}

\emailAdd{fucj@ahnu.edu.cn}
\emailAdd{iascchao@ust.hk}

\abstract{
	We investigate the amplification of curvature perturbations in a two-field inflation model featuring a Gaussian potential bump. When the inflaton encounters a potential bump along the inflationary trajectory, its rolling speed is generally reduced, potentially causing a violation of the slow-roll condition. Consequently, the original decaying modes of comoving curvature perturbations during the slow-roll phase start growing, and lead to enhanced small-scale density perturbations which can produce amounts of primordial black holes (PBHs) and associated scalar-induced gravitational waves. In addition, inflaton also undergoes sudden turnings at the encounter of the Gaussian potential bump, which is insignificant to the overall curvature power spectrum due to the short duration of these turns. Our paper offers a simple example of the extension of a bump-like potential for PBH formation in a single-field inflation to a two-field case, which helps alleviate the fine-tuning of initial conditions to some extent. 
}	
\keywords{primordial black holes, ultra-slow roll, sudden turning, scalar-induced gravitational waves}

\arxivnumber{2211.11387}

\begin{document}
	
\maketitle

%%%%%%%%%%%%%%%%%%%%%%%%%%%%%%%%%%%%%%%%%%%%
\section{Introduction}

As a type of potentially existing compact object in the early Universe, the primordial black holes (PBHs) have been attracting a lot of attention over years, owing to their relevance with massive astronomical and cosmological observational phenomena \cite{Khlopov:2008qy, Carr:2020xqk, Green:2020jor, Escriva:2022duf}. 
Throughout the past few decades, massive inflationary models have been used to study PBH production, with the goal of enhancing small-scale primordial curvature disturbances while maintaining the scale-invariant spectrum seen in the cosmic microwave background (CMB). 
To the best of our knowledge, PBH formation mechanisms fall under the following broad groups: extra contributions from other fields to curvature perturbations in multi-field models \cite{Kohri:2012yw, Kawasaki:2012wr, Pi:2017gih, Anguelova:2020nzl, Palma:2020ejf, Fumagalli:2020adf, Braglia:2020eai, Braglia:2020taf, Liu:2021rgq, Cai:2021wzd, Pi:2021dft, ZhengRuiFeng:2021zoz, Kawai:2022emp, Ashoorioon:2022raz}, the non-attractor evolutions lead to the amplification of perturbations (e.g., in the presence of ultra-slow-roll (USR) phases) \cite{Garcia-Bellido:2017mdw, Germani:2017bcs, Byrnes:2018txb, Passaglia:2018ixg, Fu:2019ttf, Fu:2019vqc, Liu:2020oqe, Fu:2020lob, Inomata:2021tpx, Tasinato:2020vdk, Ragavendra:2020sop, Kawai:2021edk, Hooshangi:2022lao, Cole:2022xqc, Karam:2022nym, Fu:2022ypp, Balaji:2022rsy,Pi:2022zxs}, parametric resonance or tachyonic instability of curvature perturbations \cite{Cai:2018tuh, Dimopoulos:2019wew, Chen:2019zza,Cai:2019bmk, Ashoorioon:2019xqc, Chen:2020uhe, Zhou:2020kkf, Peng:2021zon, Addazi:2022ukh}, the non-Gaussianity of distribution probability \cite{Ezquiaga:2019ftu, Atal:2019erb, Figueroa:2020jkf, Cai:2021zsp, Cai:2022erk, Matsubara:2022nbr, Ferrante:2022mui} and the growth of perturbations in a contracting universe \cite{Chen:2016kjx, Quintin:2016qro, Chen:2022usd}.

Recently, it has been noticed in Ref. \cite{Mishra:2019pzq} that inflaton's rolling speed can be slowed down when it climbs over a bump-like potential or passes a dip-like potential, amplifying comoving curvature perturbations. Similar models involving a step-like potential have been investigated in the literature \cite{Inomata:2021uqj,Inomata:2021tpx,Cai:2021zsp, Cai:2022erk}.
Reference \cite{Inomata:2021tpx} pointed out that the amplification of perturbations can be simply explained by the energy conservation, such that whether inflation climbs up or down the potential, the energy transfer between its kinematic and potential energies causes rapid conservation of the incoming positive modes of comoving curvature perturbations into a superposition of the positive and negative modes, implying the particle productions. 
More intriguingly, Refs. \cite{Cai:2021zsp, Cai:2022erk} revealed non-perturbative effects on the tail of probability distribution of curvature perturbations even from a tiny step on inflation's potential, which may have a significant impact on the PBH abundance.
Inspired by these works, we investigate the possibility of enhancing comoving curvature perturbations in terms of a two-dimensional potential bump $V_\text{bump}(\phi, \chi)$ on the top of a basis potential $V_\text{basis}(\phi, \chi)$ in a two-field scenario, instead of single-field inflation models considered in all abovementioned literature \cite{Mishra:2019pzq, Inomata:2021uqj, Inomata:2021tpx, Cai:2021zsp, Cai:2022erk}. The main advantage of using a two-field model is that the inflaton's rolling speed can be naturally reduced as it encounters a two-dimensional potential barrier along its trajectory, and the inflaton can easily round this barrier, rather than having to fine-tune the initial condition to ensure that the inflaton can pass the bump in a single-field case \cite{Mishra:2019pzq, Inomata:2021tpx}. To demonstrate this possibility concretely, we generalize the single-field Gaussian bump studied in Ref.~\cite{Mishra:2019pzq} to a two-dimensional Gaussian bump, allowing us to realize the enhanced comoving curvature spectrum. Moreover, the amplification of comoving curvature perturbation in terms of a non-Gaussian bump has also been discussed in Appendix \ref{app:ng}.

\begin{figure}[ht]
	\centering
	\includegraphics[width=0.35\textheight]{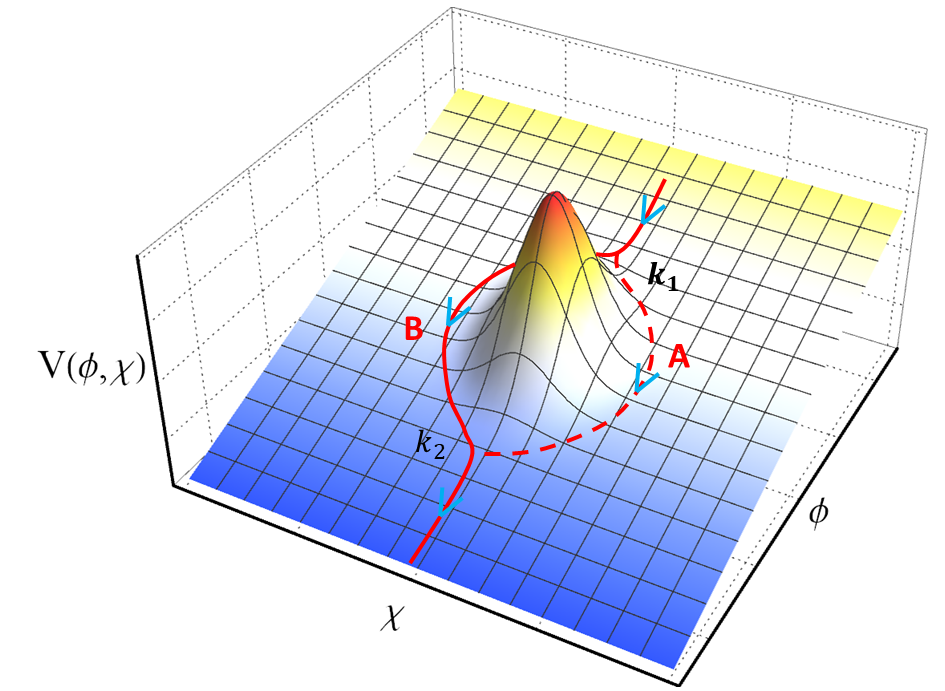}
	\caption{An illustration of the multi-stream inflation scenario. At the encounter of a potential barrier, i.e., at the scale $k_1$, Inflaton's trajectory bifurcates into two trajectories $A$ and $B$, which then converge into a single trajectory at the scale $k_2$.}
	\label{fig:ms}
\end{figure}

In the multi-field case, the presence of a bump-like potential can also result in multiply inflationary trajectories in the multi-field case, which is known as the multi-stream inflation and was first proposed in Ref. \cite{Li:2009sp}. The inflaton will fall along various trajectories with associated probabilities, resulting in multiple patches in the Universe. 
The local physics within patches may differ from one another, based on the local shape of the potential along each trajectory. As a result, the multi-stream inflation scenario predicts several intriguing observational phenomena: (i) The trajectories typically have different e-folding numbers, and their difference $\Delta N$ contributes to the extra curvature perturbations \cite{Li:2009sp, Wang:2010rs}. These additional curvature perturbations can explain the CMB cold spot \cite{Afshordi:2010wn} or trigger the formation of PBHs \cite{Nakama:2016kfq}, depending on the scale of a bifurcation; (ii) Physical environments may differ along trajectories, which can account for the spectrum asymmetry in the CMB \cite{Li:2009sp, Wang:2013vxa}, the initial cluster of PBHs \cite{Ding:2019tjk}, the stellar bubbles \cite{Cai:2021zxo}, the cosmological timer \cite{Cai:2021fgm}, and reconcile cosmic dipolar tensions as well \cite{Cai:2022dov}; (iii) The domain wall in between $A$ and $B$ may have observational effects \cite{Li:2009me}, and the thickness of the domain wall can help evade the Sunyaev-Zeldovich effect constraint, relieving the Hubble tension \cite{Ding:2019mmw}. 
According to the lattice simulation in Ref. \cite{Cai:2021hik}, the temporary domain wall with time-varying tension will be produced from the oscillatory lower-probability trajectory.

A shifted two-dimensional Gaussian bump is considered in this paper.
Because the symmetry is broken in one field direction (i.e., the $\chi$-direction shown in Fig. \ref{fig:ms}), the inflaton will almost certainly follow a single trajectory (the probability of another trajectory is exponentially suppressed \cite{Cai:2021hik}), and the multiply streams become a ``single stream''. 
In this model, we find that inflaton goes through several quasi-constant-roll (quasi-CR) phases, during which the ordinary ``decaying modes'' of comoving curvature perturbations begin to grow and give rise to the enhanced spectrum at the end of inflation.
Several resulting features of comoving curvature perturbations can be explained by the matching method studied for USR inflation \cite{Byrnes:2018txb, Carrilho:2019oqg}. 
Furthermore, inflaton makes abrupt turns when it falls off from the Gaussian potential, resulting in a tachyonic instability of isocurvature modes that transfer to adiabatic modes via the interaction.
Since the duration of each sudden turning is very short, this type of contribution to the overall curvature spectrum is negligible compared to quasi-CR phases. 
This enhanced curvature spectrum can produce a large number of PBHs which can explain the LIGO/Virgo GW events, as well as the scalar-induced gravitational waves (SIGWs) can account for the NANOGrav's suspected signal of stochastic gravitational wave background \cite{NANOGrav:2020bcs}.

This paper is organized as follows. In Sec. \ref{sec:model}, we present the model's construction details as well as the dynamics of background and perturbations in a generic setting. 
Then, using matching calculations of five successive CR phases and the sudden-turning phase, we make analytical estimates of the comoving curvature perturbations, which are consistent with the numerical results.
The PBH abundance is calculated in Sec. \ref{sec:pbh_igw} using the enhanced curvature perturbations derived in Sec. \ref{sec:model}, which is permitted by the current PBH constraints. 
The associated SIGWs turn out to be likely detectable by the current and future multi-frequency GW experiments. 
Finally, we summarize the results in Sec. \ref{sec:conclusion}.

\begin{figure}[ht!]
	\centering
	\includegraphics[width=0.28\textheight]{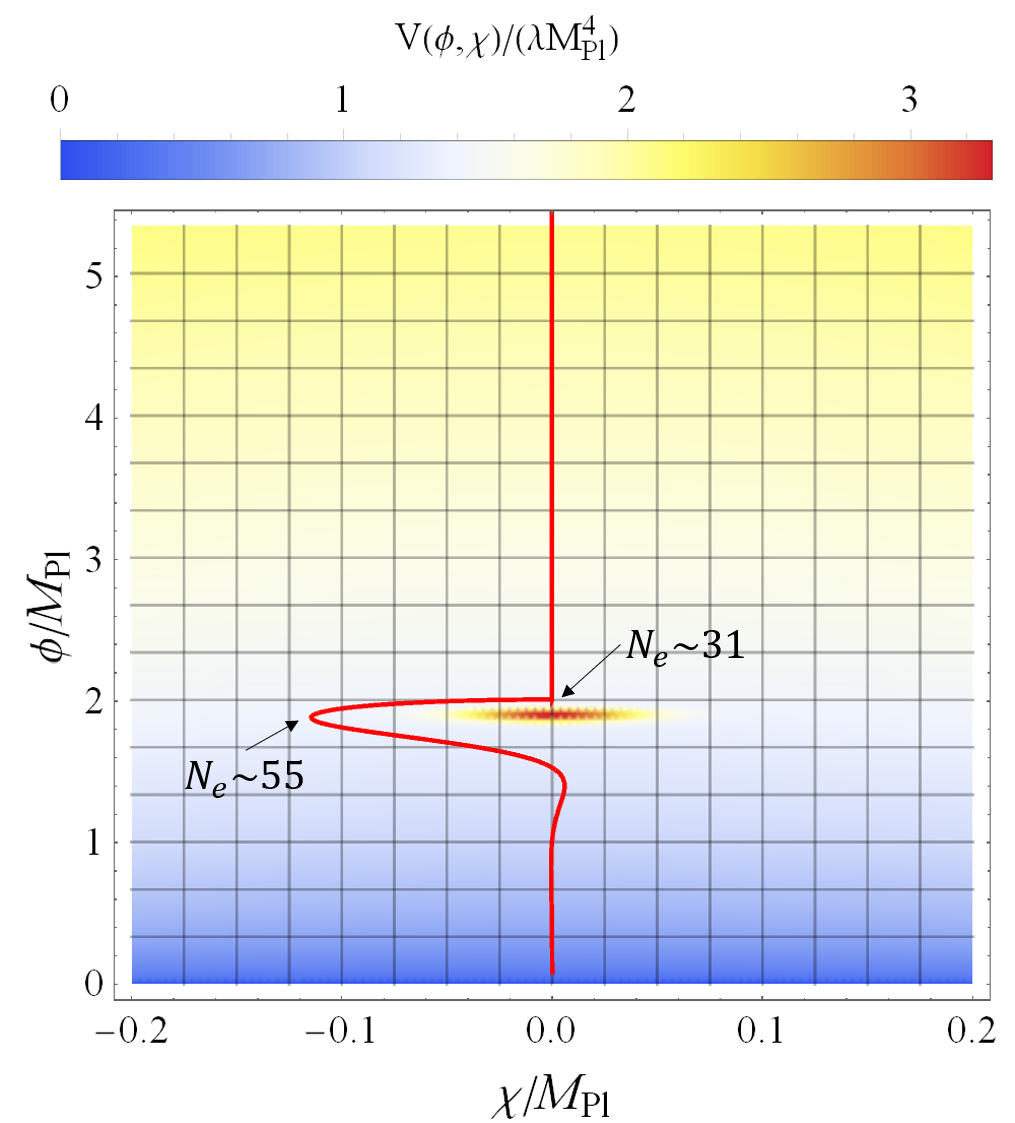}	
	\includegraphics[width=0.28\textheight]{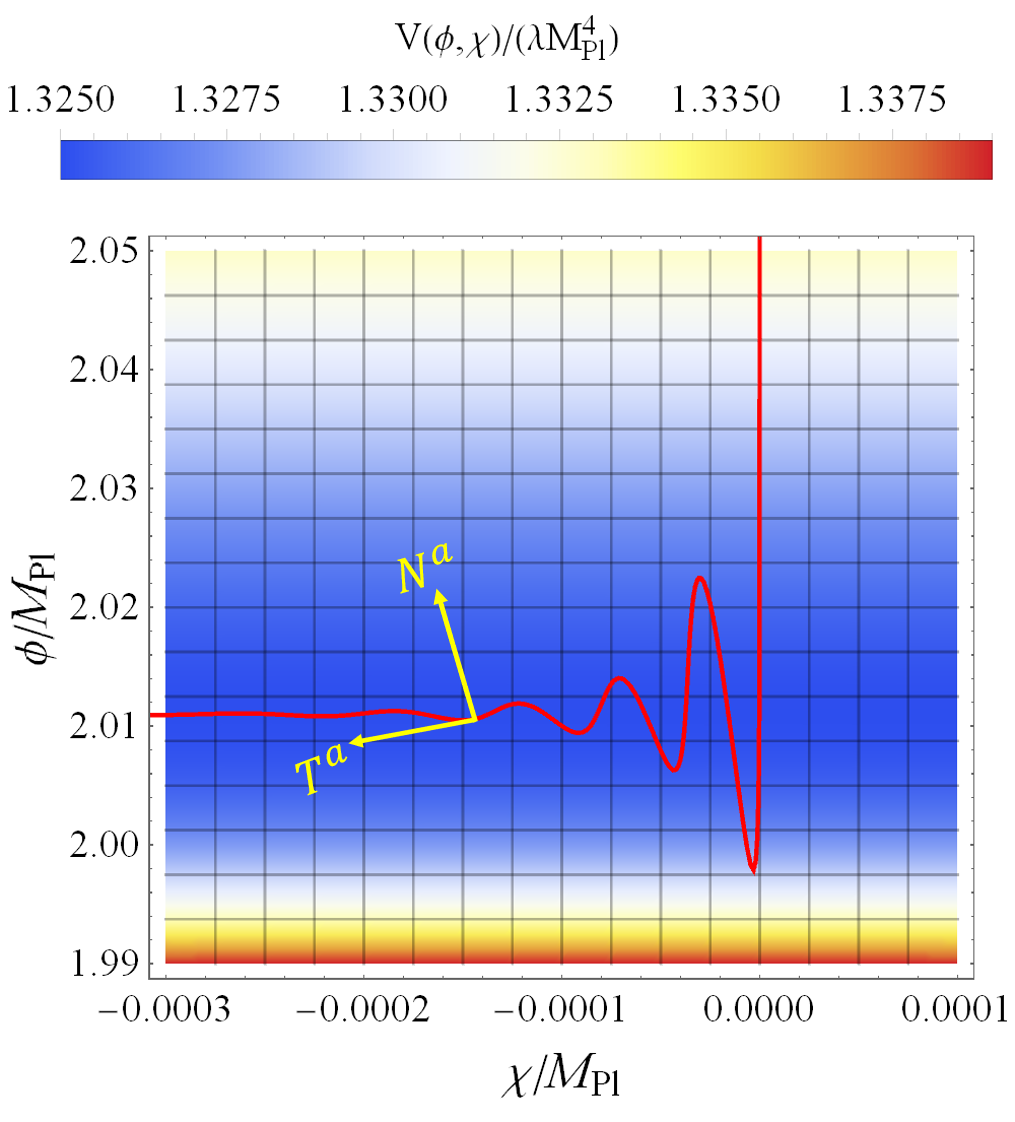}	
	\includegraphics[width=0.28\textheight]{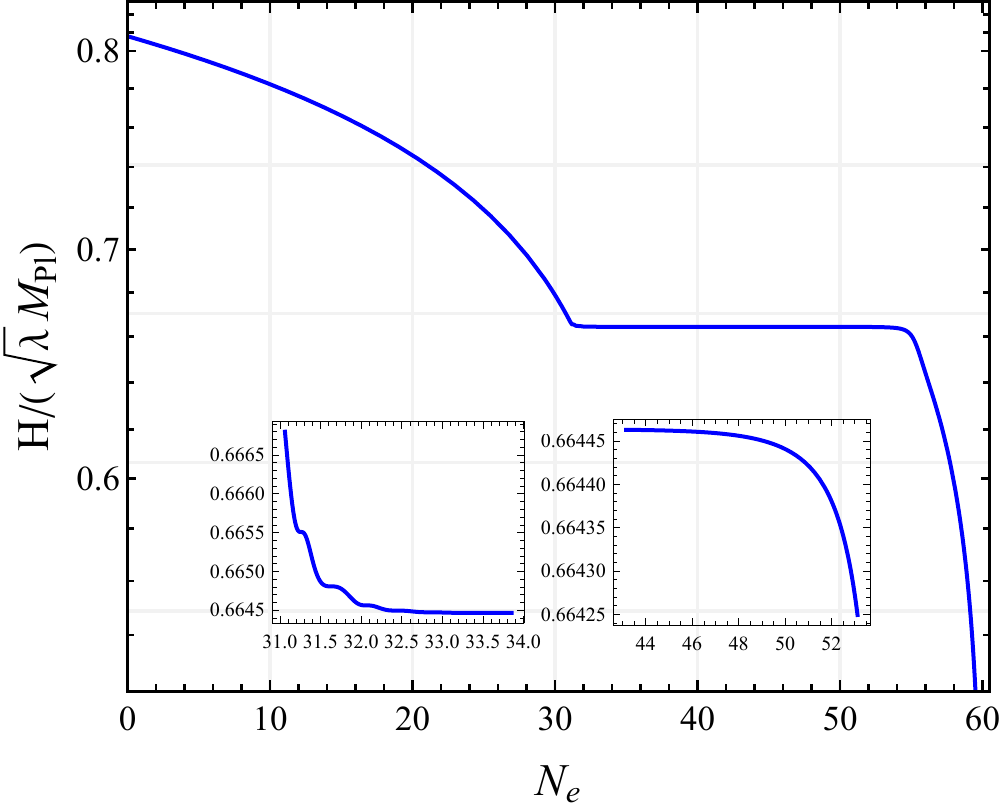}
	\includegraphics[width=0.28\textheight]{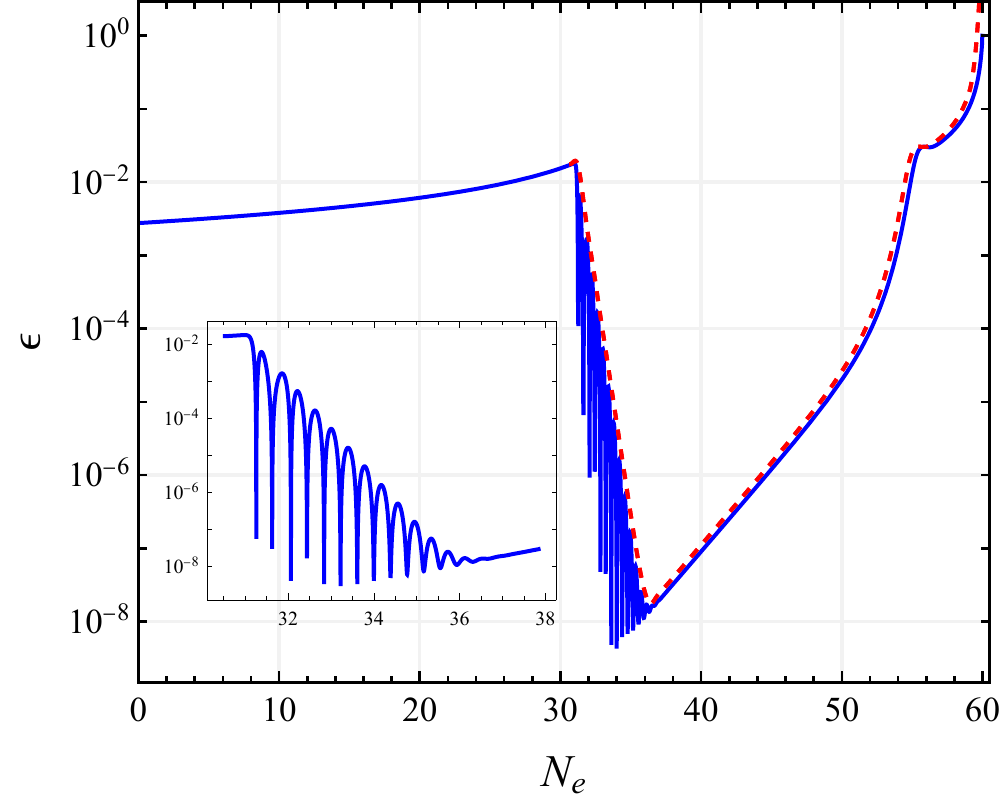}
	\includegraphics[width=0.28\textheight]{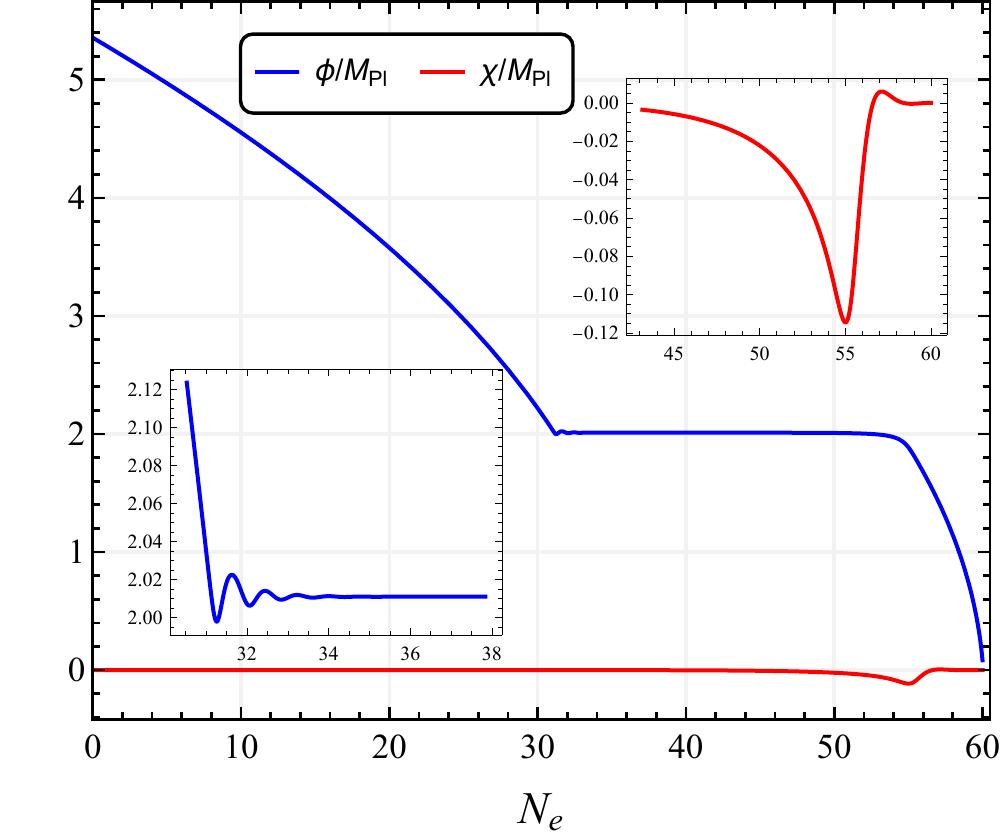}
	\includegraphics[width=0.28\textheight]{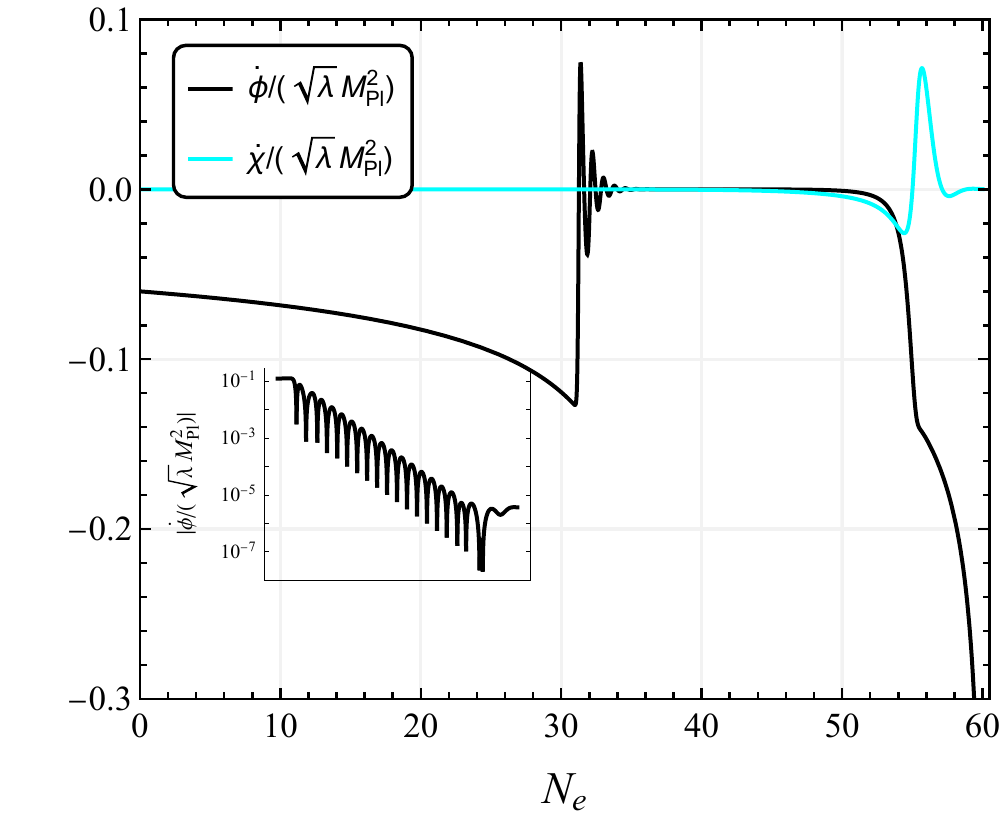}
	\caption{The background evolutions of our model \eqref{eq:action} and \eqref{eq:potential}.
		{\it Top-left panel:} The red curve denotes the inflationary trajectory in the presence of the Gaussian potential, while the shaded region corresponds to the magnitude of the potential \eqref{eq:potential}. The presence of sudden turnings is clearly shown in the {\it top-right panel}, whereas the quasi-CR phases are exhibited in the {\it middle-left \& -right panels} for the Hubble parameter and SR parameter, respectively. The red dashed curve represents the smoothed SR parameter used for making analytical estimates through matching calculations of quasi-CR phases.
		{\it Bottom-left \& -right panels} show the evolutions of background fields and their respective speeds during the whole inflationary stage.
		The values of parameters are taken as the parameter set 1 listed in Table \ref{tab:parameter}. }
	\label{fig:bg}	
\end{figure}

%%%%%%%%%%%%%%%%%%%%%%%%%%%%%%%%%%%%%%%%%%%%%%%%%%%%%%%%%%%%%%%%%%%%%%%%%%%
\section{Sudden braking and turning with a Gaussian potential bump}
\label{sec:model}

\subsection{Dynamics of background and perturbation}

We consider a simple concrete model of the multi-stream inflation scenario, which consists of two canonical scalar fields minimally coupled to Einstein gravity,
\be \label{eq:action}
S = \int \mathrm{d}^4x \sqrt{-g} \l[ {M_\text{Pl}^2 \over 2} R - {1\over2} \nabla_\mu\phi \nabla^\mu\phi - {1\over2} \nabla_\mu\chi \nabla^\mu\chi - V(\phi, \chi) \r] ~,
\ee
where $M_\text{Pl}$ is the reduced Planck mass and $R$ is the Ricci scalar built from the spacetime metric $g_{\mu\nu}$. We consider a Gaussian bump on the top of a basis potential, namely $V(\phi, \chi) = V_\text{basis}(\phi, \chi) + V_\text{bump}(\phi, \chi)$ where
\be \label{eq:potential}
V_\text{basis}(\phi, \chi) = \lambda M_{\text{Pl}}^{18/5}  \phi^{2/5} + {1\over2} m^2 \chi^2 ~,
\quad
V_\text{bump}(\phi, \chi) = \Lambda^4 \exp\l[ - {(\phi - \phi_c)^2 \over 2 \sigma_\phi^2} - {(\chi - \chi_c)^2 \over 2 \sigma_\chi^2} \r] ~,
\ee
and the field $\phi$ has a power-law potential with the fractional power $2/5$ \cite{Silverstein:2008sg}.
Here, the dimensionless parameter $\lambda$ controls the mass of the field $\phi$ and $m$ is the bare mass of the field $\chi$. The constant $\Lambda$ with the dimension of mass controls the height of the Gaussian bump, $\phi_c$ and $\chi_c$ control the peak position of potential, while $\sigma_\phi$ and $\sigma_\chi$ determine its width.

The dynamics of double fields can be deduced from the action \eqref{eq:action} with the potential \eqref{eq:potential}. Decomposing the fields into the backgrounds and perturbations: $\phi(t, \mathbf{x}) = \phi_0(t) + \delta\phi(t, \mathbf{x})$, $\chi(t, \mathbf{x}) = \chi_0(t) + \delta\chi(t, \mathbf{x})$, the EoMs for background fields in the spatially flat Friedmann-Lema\^{\i}tre-Robertson-Walker (FLRW) metric are given by, 
\bl \label{eq:bg_phi}
\ddot{\phi}_0 + 3 H \dot{\phi}_0 + {2\over5} \lambda M_{\text{Pl}}^{18/5} \phi^{-3/5}_0 - {\phi_0 - \phi_c \over \sigma_\phi^2} \Lambda^4 \exp\l[ - {(\phi_0 - \phi_c)^2 \over 2 \sigma_\phi^2} - {(\chi_0 - \chi_c)^2 \over 2 \sigma_\chi^2} \r] &= 0 ~,
\\ \label{eq:bg_chi}
\ddot{\chi}_0 + 3 H \dot{\chi}_0 + m^2 \chi_0 - {\chi_0 - \chi_c \over \sigma_\chi^2} \Lambda^4 \exp\l[ - {(\phi_0 - \phi_c)^2 \over 2 \sigma_\phi^2} - {(\chi_0 - \chi_c)^2 \over 2 \sigma_\chi^2} \r] &= 0 ~,
\el
where $H = \dot{a} / a$ is the Hubble parameter, and the Firedmann equation gives
\be \label{eq:hubble}
H^2 = {1 \over 3 M_{\text{Pl}}^2} \l[ \frac12 \dot{\phi}_0^2 + \frac12 \dot{\chi}_0^2 + V(\phi_0, \chi_0) \r] ~.
\ee
The above background equations \eqref{eq:bg_phi}, \eqref{eq:bg_chi} and \eqref{eq:hubble} determine the trajectory of inflaton in the field space spanned by $\{\phi, \chi\}$, as shown in the top-left panel of Fig. \ref{fig:bg}. 
It is convenient to apply the covariant formalism to the multifield inflation \cite{Gordon:2000hv,GrootNibbelink:2001qt, Achucarro:2010da, Achucarro:2012yr, Wang:2019gok}, which treats multifield perturbations systematically and geometrically (see Refs. \cite{Gordon:2000hv,GrootNibbelink:2001qt, Achucarro:2010da, Achucarro:2012yr, Wang:2019gok} for the detailed discussions). One decomposes field perturbations $Q_I \equiv \delta\phi_I + \frac{\dot{\phi}_I}{H} \psi$, where $\phi_{I=1,2} = \{\phi_0, \chi_0\}$ and $\delta\phi_{I=1,2} = \{\delta\phi, \delta\chi\}$, along the tangent and perpendicular directions of the inflaton's trajectory, and defines the comoving curvature (adiabatic) and entropy perturbations as
\be \label{eq:def_RF}
\mathcal{R} \equiv { H \over \dot{\varphi}_0 } T_a Q^a = { H \over \dot{\varphi}_0 } T_a \delta\phi^a ~, 
\quad
\mathcal{F} \equiv N_a Q^a = N_a \delta\phi^a ~,
\ee
respectively. Here $\dot{\varphi}_0 \equiv \sqrt{\dot{\phi}_0^2 + \dot{\chi}_0^2 }$. Note that we set $\psi = 0$ for the comoving curvature perturbation $\mathcal{R}$. The unit tangent and normal vectors are given by $T^a = { 1 \over \dot{\varphi}_0 } ( \dot{\phi}_0, \dot{\chi}_0 )$ and $N^a = { 1 \over \dot{\varphi}_0 } ( -\dot{\chi}_0, \dot{\phi}_0 )$, which satisfy the orthogonality and normalization: $T^a N_a = 0$, $T^a T_a = N^a N_a = 1$, as illustrated in the top-right panel of Fig. \ref{fig:bg}. The Latin indices are raised or lowered by the Kroneck delta function. It is customary to define two dimensionless slow-roll (SR) parameters $\epsilon$ and $\eta^a$ to account for the background dynamics of inflation, namely $\epsilon \equiv - { \dot{H}  \over H^2 } = { \dot{\varphi}_0^2 \over  2 M_{\text{Pl}}^2  H^2 }$ and $\eta^a \equiv - { 1  \over  H \dot{\varphi}_0 } { D \dot{\phi}_0^a  \over \mathrm{d}t }$, where we introduce the covariant derivative $D_t \equiv \dot{\phi}_0^a \nabla_a$ in the field space with respect to the cosmic time $t$. The vector $\eta^a$ can tell us how quickly $\dot{\phi}_0^a$ changes over time, which can be decomposed along the tangent and normal directions as $\eta^a = \eta_\| T^a + \eta_\perp N^a$ and we define $\eta_\| \equiv - { \ddot{\varphi}_0 \over H \dot{\varphi}_0 }$ and $\eta_\perp \equiv { V_N \over H \dot{\varphi}_0 }$, where $\eta_\|$ is recognized as the usual SR parameter in the single-field inflation, and $\eta_\perp$ indicates how fast $T^a$ rotates and thus describes the turning rate of the trajectory $\dot{\phi}_0^a$. The quadratic action for curvature and entropy perturbations can be derived from the action \eqref{eq:action} as \cite{Achucarro:2010da, Achucarro:2012yr}
\be \label{action_RF}
S^{(2)} =
\int \mathrm{d}t \mathrm{d}^3\mathbf{x} a^3 \l[ 
M_{\text{Pl}}^2 \epsilon \dot{ \mathcal{R} }^2 
- M_{\text{Pl}}^2 \epsilon { (\nabla \mathcal{R})^2 \over a^2}
+ 2 \sqrt{2 \epsilon}  M_{\text{Pl}} \Omega \dot{\mathcal{R}} \mathcal{F}
+ \frac12 \dot{\mathcal{F}}^2 - \frac12 { (\nabla\mathcal{F})^2 \over a^2 } 
- \frac12 M^2 \mathcal{F}^2  \r] ~,
\ee
where we have defined the entropic mass of the entropy field $\mathcal{F}$ as
\be \label{entropy_mass}
M^2 = V_{NN} + \epsilon H^2 M_{\text{Pl}}^2 \mathbb{R} - \Omega^2 ~,
\ee
where $V_{NN} \equiv N^a N^b \nabla_a \nabla_b V$ and $\mathbb{R}$ is the Ricci scalar of the field space. The quantity $\Omega \equiv H \eta_\perp$ is the turning rate describing the bends of a trajectory. It is clear seen from Eq. \eqref{entropy_mass} that the entropic mass is also affected by the turning rate of the trajectory, since the potential receives a correction coming from the centripetal force experienced by the turning. In our model \eqref{eq:action}, since the geometry of field space is trivial, the ``bare" entropic mass is merely contributed by the potential, namely $m_\text{bare}^2 \equiv V_{NN} + \epsilon H^2 M_{\text{Pl}}^2 \mathbb{R} = V_{NN}$. 
Notice that at the quadratic level, the interaction between adiabatic and entropic modes is controlled by a dimensionless quantity $\eta_\perp$.
From the action \eqref{action_RF}, we derive coupled dynamical equations for curvature and entropy perturbations,
\bl \label{eom_R}
\ddot{\mathcal{R}} + \l( 3 + 2 \epsilon - 2 \eta_\| \r) H \dot{\mathcal{R}} + \frac{k^2}{a^2} \mathcal{R}
&= - 2 \Omega \frac{H}{\dot{\varphi}_0} \l[ \dot{\mathcal{F}} + \l( 3 - \eta_\| - \epsilon + {\dot{\Omega} \over H \Omega} \r) H \mathcal{F} \r] ~,
\\ \label{eom_F}
\ddot{\mathcal{F}} + 3 H \dot{\mathcal{F}} + \l( \frac{k^2}{a^2} + M^2 \r) \mathcal{F} &= 2 \Omega \frac{\dot{\varphi}_0}{H} \dot{\mathcal{R}} ~,
\el
In practice, these two coupled equations are commonly calculated numerically. However, on the superhorizon scales, they can be simplified and some estimates for their evolutions can be made.

\subsection{Anatomy of curvature and entropy perturbations}

It is evident from the top-left panel of Fig.~\ref{fig:bg} that the dynamics of our model can be categorized into three phases: 
(a) {\it Single-field phase}. Initially, the inflaton rolls down the potential along the $\phi$-direction while strapped to the local minimum on $\chi$-direction (i.e., $\chi=0$). Throughout this phase, the dynamics is entirely described by $\phi^{2/5}$ inflation model \cite{Silverstein:2008sg}, which is consistent with the Planck experiment \cite{Planck:2018jri} for approximately $30\sim40$ e-folds; 
(b) {\it Braking-Turning phase}. As inflaton encounters the shifted Gaussian barrier with a SR speed, it will ascend the barrier with a rapidly-diminishing speed $\dot{\phi}$ and eventually attain the maximum potential point when inflaton's speed is nearly zero (The corresponding parameters are calculated as $N_e \simeq 31.26$, $\phi \simeq 1.998$, $\chi \simeq -3.27 \times 10^{-6}$, $\dot{\phi} \simeq - 4.97 \times 10^{-7}$, $\dot{\chi} \simeq - 3.49 \times 10^{-5}$ and $V/(\lambda M_\text{Pl}^4) \simeq 1.33$ in terms of the parameter set 1 as listed in Table \ref{tab:parameter}). 
And then, the inflation starts entering the braking-turning phase, which involves the quasi-CR (i.e., the second SR parameter $\eta \equiv \dot{\epsilon}/(\epsilon H)$ is constant) and sudden-turning processes. During this phase, the SR parameter decreases rapidly, while the Hubble parameter remains nearly constant as indicated in the middle-right panel of Fig.~\ref{fig:bg}. 
Meanwhile, due to the asymmetry of the shifted Gaussian potential along the $\chi$-direction, the inflaton will acquire a negligible yet gradually-growing speed $\dot{\chi}$, as illustrated by the cyan curve in the bottom-right panel of Fig.~\ref{fig:bg}. 
After attaining the maximum point on the Gaussian potential, the inflationary trajectory undergoes a significant bend, followed by oscillations along the periphery of the Gaussian potential. During this phase, the true inflation direction slowly rotates towards the $\chi$-direction. The inflaton can round this Gaussian potential and the braking-turning phase ends. 
Subsequently, there is a second (c) {\it single-field phase}, in which the inflationary trajectory returns to the $\phi$-direction, and ends inflation when the SR parameter $\epsilon$ reaches unity. 
As demonstrated in Fig.~\ref{fig:sp}, for the two single-field phases (a) and (c) featuring a $\phi^{2/5}$ potential, the curvature perturbation is anticipated to be almost scale-invariant and in agreement with the Planck data. 
The primary focus of our subsequent analysis will be on the braking-turning phase (b), which constitutes the main outcome of this study.

\subsubsection{Braking phase}

The concept of USR inflation was initially explored in Ref. \cite{Kinney:2005vj} (with similar topics being discussed in Refs. \cite{Seto:1999jc, Inoue:2001zt}), while the CR inflation was first investigated in Ref. \cite{Motohashi:2014ppa}. 
The CR and USR phases are generally non-attractor\footnote{The conditions for the existence of an attractor behavior during the USR phase have been explored in Refs.~\cite{Salvio:2017oyf, Pattison:2018bct}.} and therefore exhibit distinct features relative to the standard SR phase. Specially, CR or USR is a single-field inflation but not a single-clock inflation, since the decaying mode of curvature perturbations could surpass the constant mode. Consequently, the long-wavelength curvature perturbations cannot be treated as a local time reparametrization of background. 
To be more precise, the long-wavelength comoving curvature perturbations evolve as $\mathcal{R}_k \simeq C_k + D_k \int^\tau {d\tau' \over a^2 \epsilon}$, and thus the second term (known as the ``decaying mode'') ceases to decay once $\eta \leq -3$. 
The direct outcome of this decaying mode is the growth of superhorizon curvature perturbations and the ``steepest'' sustained growth rate\footnote{The ``sustained'' here means that such special growth can persist over multiple e-folds. The transient super-$k^4$ growth was reported in Refs. \cite{Ozsoy:2019lyy, Tasinato:2020vdk}.} $k^4$ of the curvature power spectrum in single-field inflation has been discovered in Ref. \cite{Byrnes:2018txb}. 
If an additional CR phase with $\eta=-1$ is included between the SR and USR phases, a more rapid sustained growth rate $k^5 (\ln k)^2$ can be observed \cite{Carrilho:2019oqg}. 
Recently, Ref. \cite{Cole:2022xqc} has revealed that the mass function of PBHs is insensitive to the steepness of the power spectrum when it is steeper than $k^2$. However, other relevant detections for PBHs (e.g., via SIGWs) may be influenced by the shape of the curvature spectrum which is needed to investigate further. Furthermore, Ref.~\cite{Cole:2022xqc} has identified several artifacts arose from unphysical instantaneous transitions between SR, CR and USR phases that are commonly assumed in the literature. For example, the reduction of oscillations and the peak amplitude of the curvature power spectrum, with the latter having a significant impact on the PBH mass function. Moreover, the super-$k^4$ sustained growth may be erased by considering more realistic smooth transitions.

\begin{figure}[ht!]
	\centering 
	\includegraphics[width=0.33\textheight]{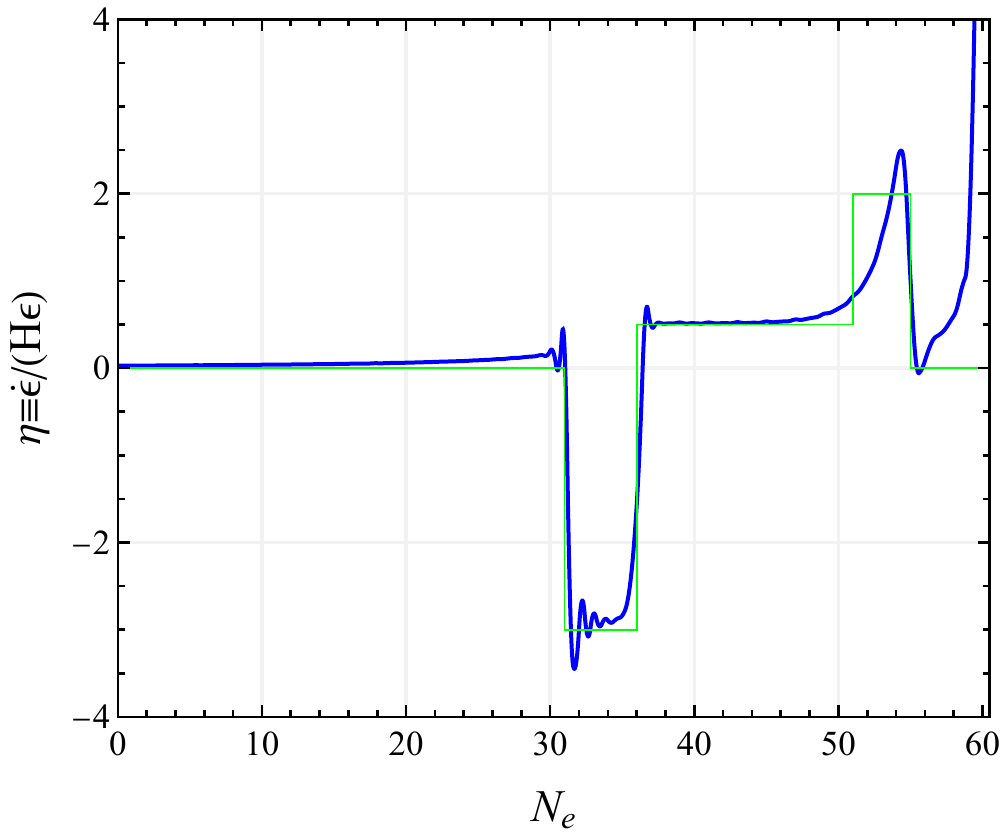}	
	\includegraphics[width=0.342\textheight]{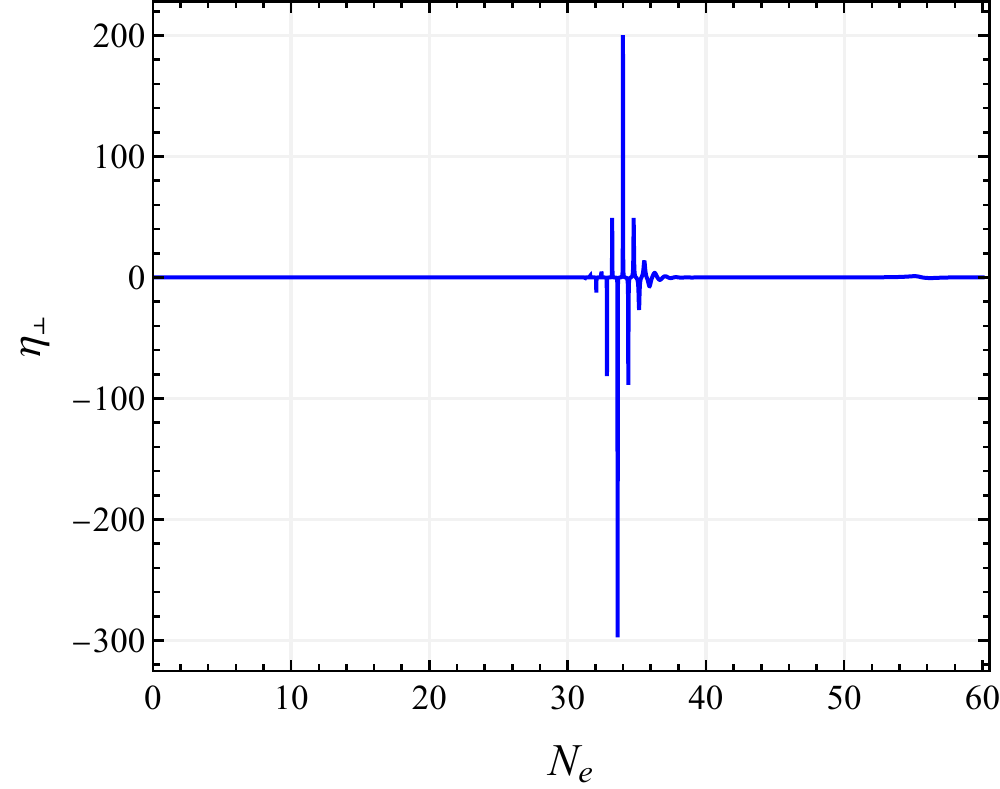}
	\caption{{\it Left panel:} The second SR parameter $\eta$ (blue) is obtained from the smoothed SR parameter shown by the red dashed curve in the middle-right panel of Fig. \ref{fig:bg}. We approximate $\eta$ by five successive CR phases $\eta = 0 \rightarrow \eta =-3 \rightarrow \eta =0.5 \rightarrow \eta =2 \rightarrow \eta = 0$ with instant transitions (green).
		{\it Right panel:} The dimensionless parameter $\eta_\perp$ shows several sudden turnings during the encounter of the Gaussian potential. In spite of high spikes of $\eta_\perp$, they do not affect the overall perturbations significantly, since the short durations of turnings.
		The values of parameters are taken as the parameter set 1 listed in Table \ref{tab:parameter}.}
	\label{fig:bg2}
\end{figure}

In our case, the quasi-CR phases occurred when the inflationary trajectory was primarily aligned along one field direction (either $\phi$ or $\chi$), as shown in the bottom panel of Fig. \ref{fig:bg}. Hence, we can apply similar analytical calculations for the USR phase in the single-field inflation to our model. 
In what follows, we adopt the matching calculations introduced in Ref. \cite{Byrnes:2018txb} to estimate the curvature power spectrum by assuming instantaneous transitions for simplicity (refer to Appendix \ref{app:match} for details). To capture the essential features of this analytical approximation, we smooth the SR parameter $\epsilon$ represented by the red dashed curve in the middle-right panel of Fig. \ref{fig:bg}, and then plot the second SR parameter $\eta$ in the left panel of Fig. \ref{fig:bg2}. 
As an approximation, we incorporate five successive CR phases: $\eta = 0 ~(31) \rightarrow \eta =-3 ~(5) \rightarrow \eta =0.5 ~(15) \rightarrow \eta =2 ~(3) \rightarrow \eta = 0$, where the values in the parentheses refer to the e-folding numbers of the corresponding CR phases. 
The estimated spectrum of the comoving curvature perturbations shown by the thin green curve in Fig. \ref{fig:sp} is close to the numerical result (the black curve). This observation implies that the braking phase is the major contributor to the comoving curvature perturbations in our model.

Furthermore, the growth and decay slopes of the spectra depicted in Fig.~\ref{fig:sp} can be approximated using the methods described in Ref. \cite{Carrilho:2019oqg}.
At the end of inflation, all observable modes exit the horizon, allowing for the growth and decay slopes of the final spectrum to be approximated via the asymptotic expansion [i.e., from Eq. \eqref{rk_bd}] on the superhorizon scales in a single $\eta$-phase \cite{Carrilho:2019oqg},
\be \label{expansion}
\mathcal{R}_k \simeq A k^{-{3+\eta\over2}} + B (-\tau)^2 k^{{1-\eta\over2}} + C (-\tau)^{3+\eta} k^{{3+\eta\over2}} ~,
\ee
where $A, B, C$ are normalization constants determined by either the initial conditions or matching conditions, the $A$- or $C$-term represents the constant mode that depends on the value of $\eta$, while the $B$-term always decays during inflation. In the SR phase, $\eta \simeq 0$, the $B$-term is the subleading decaying mode and would lead to $k^4$ scaling if it is dominated after the transition. This simple asymptotic expansion shows the origin of the $k^4$ growth \cite{Carrilho:2019oqg}. 
The key point here is that, if the dominated mode in a $\eta$-phase before the transition is not constant, only this leading term will be imprinted on the solution after the transition, thereby determining the growth or decay slope of the final spectrum. 
Conversely, if the dominated mode is constant before the transition, the eventual growth or decay slope after the transition depends on the interplay between this constant mode and other modes that are undergoing growth.
For $\eta \leq -3$, the dominated mode in the $\eta$-phase before the transition is not constant (i.e., the $C$ term, identified as a growing mode in this case), and thus we can directly determine the index of the curvature power spectrum as
\be \label{index}
n_s - 1 = 3 - |3 + \eta| ~,
\ee
which is clearly to see the bound on the slope $n_s - 1 \leq 3$, and the scale-invariant spectra for $\eta=0, -6$\footnote{Note that the relation \eqref{index} holds for the normal SR case such that the $A$-term is always dominated before and after the transition, so we keep the absolute symbol in Eq. \eqref{index}.}. For $\eta > -3$, the dominated mode in the expansion \eqref{expansion} is the $A$-term, and thus the decaying modes $B$- or $C$-term would be dominated after the transition, the growth rate can be determined as $n_s - 1 = 5 - |1 + \eta|$. Evidently, for $\eta=0$, we would have $k^4$ growth after the transition, which is consistent with the above discussion. The above analyses based on the superhorizon asymptotic expansion \eqref{expansion} provide two crucial insights into curvature perturbations during the CR and USR phases: (i) the growing modes stem from subleading terms (typically decaying modes) before the transition; (ii) there is superhorizon growth of curvature perturbations. 
The growth slopes for the red, blue, black curves in Fig.~\ref{fig:sp} are approximately $n_s - 1 \simeq 3$, which are mainly governed by the interplay among three terms in the expansion \eqref{expansion} and other higher-order terms that are not shown explicitly in Eq. \eqref{expansion}. For the scales of interest\footnote{The modes exit the horizon at each transition of $\eta = 0 \rightarrow \eta = -3 \rightarrow \eta = 0.5 \rightarrow \eta = 2 \rightarrow \eta = 0$ in terms of the parameterc set 1 listed in Table \ref{tab:parameter}, are given by $(1.2 \times 10^{12}, 1.8 \times 10^{14}, 5.8 \times 10^{20}, 1.2 \times 10^{22})$ Mpc$^{-1}$.}, the decay slopes are shown to be around $n_s - 1 \simeq 0.5$ in Fig. \ref{fig:sp}, which can be obtained from the relation \eqref{index} by setting $\eta=0.5$, since before the $\eta = 2$ phase, the constant mode is dominated and the decaying mode is no longer significant after the transition.

Another typical characteristic that arises from the interplay between the constant and decaying modes in Eq. \eqref{expansion} after transitions is the emergence of a dip prior to the growth in the power spectrum. This phenomenon is depicted in Fig.~\ref{fig:sp}.
It has been shown in Ref. \cite{Byrnes:2018txb} that the dip position $k_\text{dip}$ is given by $k_\text{dip} = \tau_\text{USR}^{-3/2} k_\text{USR}$, here $\tau_\text{USR} = \tau_\text{USR,start}/\tau_\text{USR,end}$ is the duration of the USR phase, and $k_\text{USR}$ is the scale exits horizon at the onset of the USR phase. This dip structure indeed implies the following superhorizon growth before the start of the USR phase. 
In our case, the dip position does not adhere to this relationship because the growth is not solely due to the individual decaying mode shown in the expansion \eqref{expansion}, but rather, its interplay with other modes. The numerical results tell us the relation $k_\text{dip} \simeq e^{-N_\text{CR}} k_\text{USR}$, where $N_\text{CR}$ is the e-folding number of the CR phase following the first SR phase. In our case, the dip essentially comes from the cancellation mentioned above, but it is more physically transparent to understand via the ``overshoot'' argument \cite{Germani:2017bcs}. 
We have to emphasize that, in a general case, the emergence of the dip can only be explained in part by the ``overshoot'' argument \cite{Byrnes:2018txb}, and it works well in our model. Overshoot means that inflaton rolls down from the SR phase to the USR phase, its acceleration need to grow to some extent to overcome the potential gradient, so its speed and SR parameter will increase before their reduction (as shown in the middle-right panel of Fig. \ref{fig:bg}) and the resulting curvature power spectrum over that range of scales would be reduced. In order to suppress this dip, one direct way is to reduce the inflaton's speed at the transition \cite{Germani:2017bcs}, which is consistent with the result shown in Fig. \ref{fig:sp}.
If the initial position of the inflaton is closer to the Gaussian potential, resulting in a lower velocity at the transition, the corresponding dip will be more subdued.
In the USR case, the peak amplitude can be simply estimated by comparing the constant and growing modes \cite{Byrnes:2018txb, Carrilho:2019oqg}. As mentioned earlier, the growth of spectrum is not simply determined in our model, and numerical calculations are necessary to obtain $k_\text{dip}$, and then the peak amplitude is approximately to $\mathcal{P}_\mathcal{R}(k_\text{USR}) \sim \mathcal{P}_\text{CMB} e^{\alpha N_\text{CR}}$, where $\alpha \simeq 3$ is the growth rate of spectrum.

In this paper, we ignore the non-Gaussianity that may be generated in our model. In fact, one distinct feature of USR is the violation of Maldacena's non-Gaussianity consistency relation in any attractor-single-field inflation, $f_\text{NL} = 5(1 - n_s)/12$\footnote{Note that the nonlinearity parameter $f_\text{NL}$ here is defined as $\langle \mathcal{R}_{\mathbf{k}_1} \mathcal{R}_{\mathbf{k}_2} \mathcal{R}_{\mathbf{k}_3} \rangle \equiv (2\pi)^3 \delta(\mathbf{k}_1+\mathbf{k}_2+\mathbf{k}_3) {12\over5} f_\text{NL} \mathcal{P}_\mathcal{R}(k_1) \mathcal{P}_\mathcal{R}(k_3)$ in the squeezed limit $k_1 \ll k_2 =k_3$.}, which connects the squeezed limit of the bispectrum to the power spectrum \cite{Maldacena:2002vr}. Hence, the non-Gaussianity is negligible for the SR single-field inflation where $n_s \simeq 1$. In contrast, in the USR limit $\eta = -6$, the nonlinearity parameter is found to be $f_\text{NL} = 5(3 - n_s)/4$ even for any configuration \cite{Martin:2012pe}, and it becomes $5/2$ when the power spectrum is scale-invariant \cite{Namjoo:2012aa}. However, the USR phase needs to be followed by a SR phase in order to solve the horizon and flatness problems without fine-tuning of an initial condition, and explain CMB observed primordial density perturbations. 
Taking into account the transition to an attractor phase following the USR phase, Ref.~\cite{Cai:2018dkf} demonstrated that the significant non-Gaussianity produced during the USR phase can be partly or entirely eliminated by this transition phase, and Maldacena's consistency relation remains violated.
However, several discrepancies have been reported in the literature \cite{Bravo:2017wyw, Passaglia:2018ixg, Bravo:2020hde, Suyama:2021adn} regarding the non-Gaussianity associated with non-attractor phases. Moreover, the quantum stochastic effect associated with USR phase may also have an impact on the amplification of curvature perturbations \cite{Ballesteros:2020sre, Pattison:2021oen, Rigopoulos:2021nhv}. 
In light of the above mentioned uncertainties and the complexity in two-field inflation, we ignore the non-Gaussainity and quantum stochastic effect in this paper for simplicity, and focus on the growth of curvature perturbations from transitions in between different constant-$\eta$ phases.

\begin{table}[h]
	\centering
	\resizebox{\textwidth}{!}{
		\begin{tabular}{|c|c|c|c|c|c|c|c|c|c|}
			\hline
			&\multicolumn{6}{c|}{Parameter Sets} &\multicolumn{3}{c|}{Results}
			\\
			\cline{2-10}
			\multicolumn{1}{|c|}{} 
			& $m^2/(\lambda M_{\text{Pl}}^2)$ & $\Lambda^4/(\lambda M_{\text{Pl}}^4)$ & $\phi_c/M_{\text{Pl}}$ &  $\chi_c/M_{\text{Pl}}$ & $\sigma_\phi/M_{\text{Pl}}$ & $\sigma_\chi/M_{\text{Pl}}$ & $\lambda$ & $n_s$  & $r$ \\
			\hline
			\emph {1} & $2$ & $2$ & $1.90$ &	$2.63 \times 10^{-5}$ &	$0.030$ & $0.030$ & $6.89\times10^{-10}$  &	$0.9674$ &	$0.044$ 
			\\
			\emph {2}  & $2$  & $2$ & $2.73$  & $3.10 \times 10^{-5}$	& $0.024$	&$0.024$	& $7.63 \times10^{-10}$	& $0.9645$	&$0.048$
			\\
			\emph {3} & $2$	& $2$	& $3.55$	& $3.18 \times 10^{-5}$	& $0.020$ & $0.020$	& $7.80\times10^{-10}$	& $0.9639$	&  $0.049$
			\\
			\hline
		\end{tabular} 
	}
	\caption{Parameter sets in our model for a fixed peak amplitude of curvature spectra $\mathcal{P}_{\mathcal{R}} \sim 10^{-2}$ as shown in Fig. \ref{fig:sp}. The corresponding results of dimensionless parameter $\lambda$, the power index $n_s$ and the tensor-to-scalar ratio $r$ are also presented.}
	\label{tab:parameter}
\end{table}

\subsubsection{Turning phase}

\begin{figure}[h]
	\centering
	\includegraphics[width=0.45\textheight]{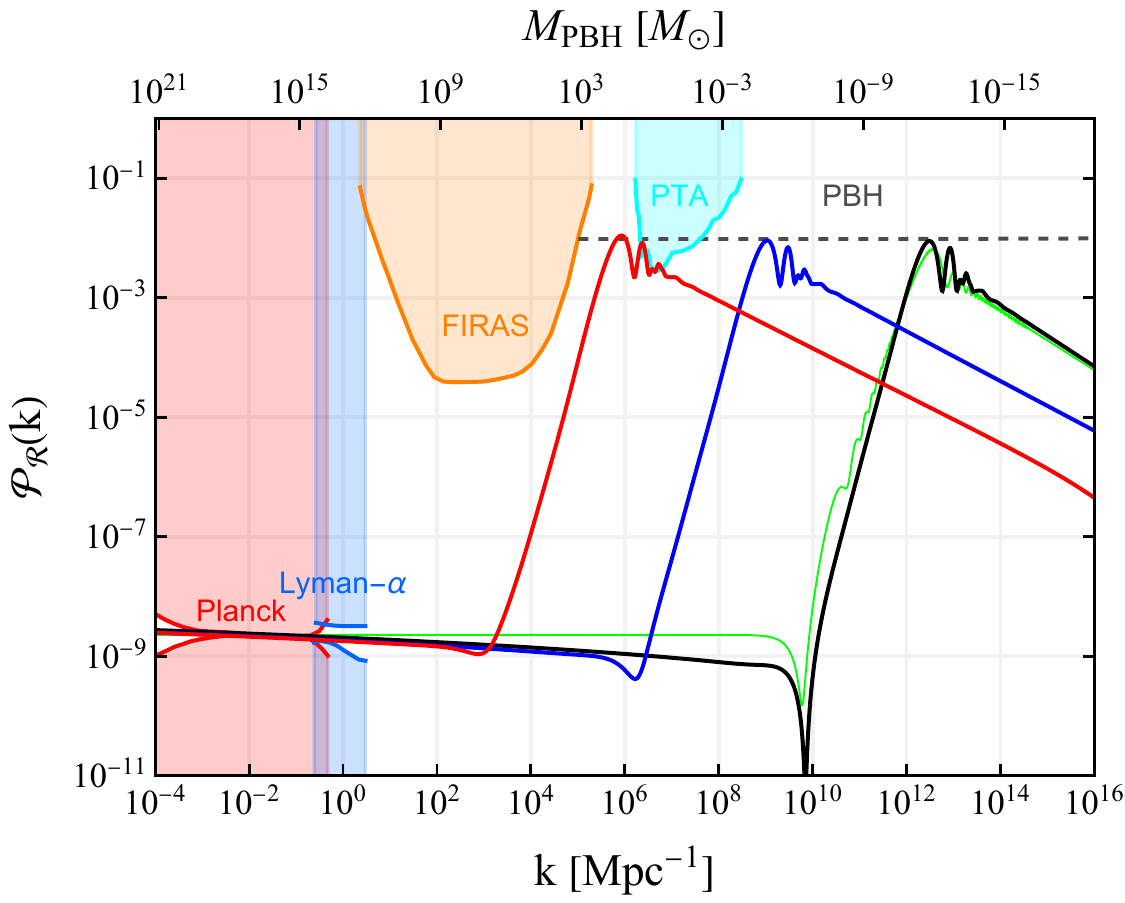}
	\caption{The numerical results of power spectra of the comoving curvature perturbations at the end of inflation. The black, blue, and red curves respectively refer to parameter sets 1-3 listed in Table \ref{tab:parameter}. The current constraints on $\mathcal{P}_\mathcal{R}(k)$ from Planck \cite{Planck:2018jri}, Lyman-$\alpha$ \cite{Bird:2010mp}, FIRAS \cite{Fixsen:1996nj} and PTA \cite{Byrnes:2018txb} are shown by the shadowed regions, while the grey dashed line refers to $\mathcal{P}_\mathcal{R} \sim 10^{-2}$ in order to produce an abundance of PBHs. 
	The thin green curve denotes the analytical results from the matching calculations.}
	\label{fig:sp}
\end{figure}

The presence of sudden turnings during $\eta\simeq-3$ phase is evident in the right panel of Fig. \ref{fig:bg2}, which corresponds to the top-right panel of Fig. \ref{fig:bg}. The strongly non-geodesic motion with $\eta_\perp^2 \gg 1$ in the field space has been investigated in extensive literature \cite{Cespedes:2012hu, Achucarro:2012sm, Noumi:2013cfa, Gao:2013ota, Garcia-Saenz:2018vqf, Garcia-Saenz:2018ifx, Achucarro:2018vey, Bjorkmo:2019fls, Garcia-Saenz:2019njm, Fumagalli:2019noh} which were primarily motivated by ultraviolet completions of the inflation scenario.
On the superhorizon scales, the master equations \eqref{eom_R} and \eqref{eom_F} can be greatly simplified as \cite{Langlois:2008mn}
\bl
\dot{\mathcal{R}}
&\simeq - 2 \Omega \frac{H}{\dot{\varphi}_0} \mathcal{F} ~,
\\
\ddot{\mathcal{F}} + 3 H \dot{\mathcal{F}} + M_\text{eff}^2 \mathcal{F} &\simeq 0 ~,
\el
where the effective mass of the entropy field is defined as $M_\text{eff}^2 \equiv M^2 + 4 \Omega^2$. It is manifest from the above two equations that on the superhorizon scales, adiabatic modes are sourced by entropy modes, while entropy modes evolve independently. A sudden turning will give rise to a negative large entropy mass square $M^2/H^2$ defined in Eq. \eqref{entropy_mass}, which signifies a transient tachyonic instability of subhorizon entropy modes $\mathcal{F}_k$ for $k^2/a^2 \lesssim |M^2|$. Through the coupling, shown in the action \eqref{action_RF}, this exponential growth of $\mathcal{F}_k$ would transfer to the adiabatic modes $\mathcal{R}_k$, resulting in exponential growth of curvature perturbations on relevant scales.
On the other hand, this sudden turning contributes a positive large effective entropy mass $M_\text{eff}^2/H^2$ on the superhorizon scales, implying a stable background and a rapid decay of superhorizon entropy modes $\mathcal{F}_k$, so that the adiabatic perturbation will remain conserved on the superhorizon scales. This fact can also be understood from the effective field theory (by integrating out the heavy entropy field) with an imaginary sound speed $c_s^2 \equiv M^2/M_\text{eff}^2$ \cite{Achucarro:2012sm, Garcia-Saenz:2018ifx, Bjorkmo:2019qno, Fumagalli:2019noh}, which is a manifestation of the exponential growth of curvature perturbations before the sound horizon crossing. One can design a turning rate to generate PBHs via the resulting enhanced curvature spectrum \cite{Palma:2020ejf, Fumagalli:2020adf, Anguelova:2020nzl, Ballesteros:2021fsp}. Hence, in our model, several sudden turnings occurred during $\eta \simeq -3$ phase will contribute the spikes around the peak position $k_\text{USR}$ of spectrum as shown in Fig. \ref{fig:sp}, however, they do not have significant contributions to spectrum on other scales.

%%%%%%%%%%%%%%%%%%%%%%%%%%%%%%%%%%%%%%%%%%%%%%%%%%%%%%%%%%%%%%%%%%%%%%%%%%%
\section{Primordial black holes and scalar-induced gravitational waves}
\label{sec:pbh_igw}
%%%%%%%%%%%%%%%%%%%%%%%%%%%%%%%%%%%%%%%%%%%%%%%%%%%%%%%%%%%%%%%%%%%%%%%%%%%

\subsection{PBH abundance}

\begin{figure}[h]
	\centering
	\includegraphics[width=0.45\textheight]{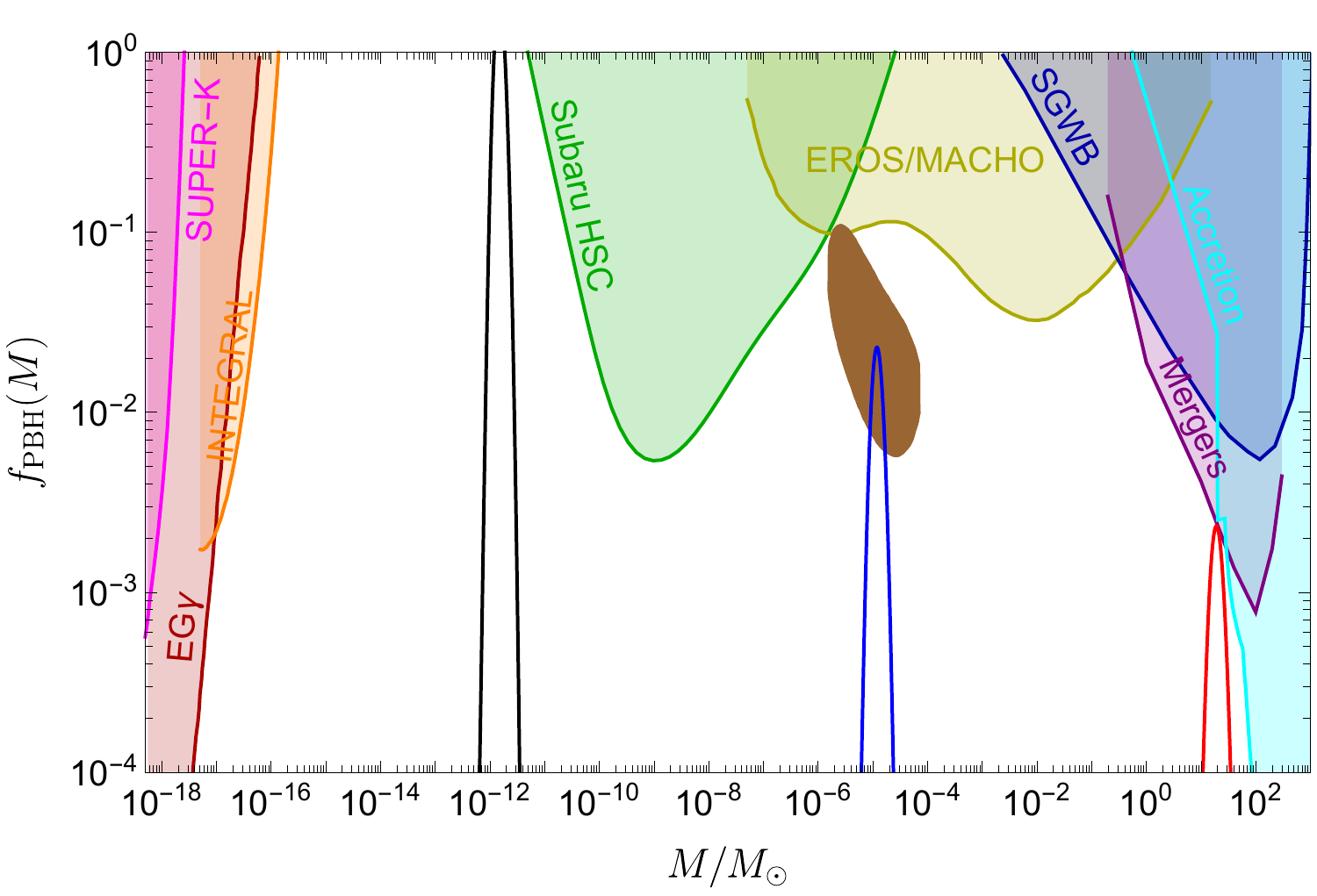}
	\caption{Three current energy fractions of PBHs $f_\text{PBH}(M)$ in our model in terms of parameter sets listed in Table \ref{tab:parameter} with various constraints on $f_\text{PBH}(M)$ adopted from Ref. \cite{Carr:2020gox}. The brown shaded region represents the PBH abundance inferred by the OGLE ultrashort-timescale microlensing events \cite{Niikura:2019kqi}.}
	\label{fig:pbh}	
\end{figure}

The large density contrast $\delta\equiv\delta\rho/\bar{\rho}$ on the small scales can cause PBH formation and the SIGWs at the horizon reentry. As previously mentioned, the comoving curvature perturbation $\mathcal{R}$ in our model is assumed to follow the Gaussian distribution, so that the same is true for the density contrast on the comoving slicing,
\be \label{delta_R}
\delta_k
= {2 \over 3} \l( {k \over aH} \r)^2 \Phi_k
\simeq {4 \over 9} \l( {k \over aH} \r)^2 \mathcal{R}_k ~,
\ee
where $\Phi$ is the Bardeen potential in the Newtonian gauge, and we ignore the anisotropic stress of matter sector for simplicity. We have utilized the relationship $\Phi \simeq {2\over3} \mathcal{R}$ on the superhorizon scales in the second equality of Eq. \eqref{delta_R}. According to the theory of gravitational instability, when the density contrast $\delta$ exceeds the threshold $\delta_c$, the overdense region will stop expansion and collapse to a BH. There have been numerous studies on the typical value of the density contrast, and the recent study \cite{Musco:2020jjb} suggests that it falls within $0.4 \lesssim \delta_c \lesssim 0.7$. Here, we take a conservative value $\delta_c = 0.6$ in this paper. The initial mass function $\beta(M)$ is calculated as
\be \label{beta_def}
\int \beta(M) \mathrm{d}\ln M \equiv {\rho_\text{PBH} \over \rho_c} ~,
\ee
where $\rho_\text{PBH}$ and $\rho_c$ are energy densities of PBHs and background at formation epoch, respectively. The most straightforward and simple method to estimate $\beta(M)$ is to use the Press-Schechter formalism \cite{Press:1973iz}, and we obtain\footnote{It should be noted that the calculation presented here actually provides the cumulative probability for the formation of a PBH with mass greater than $M$.  The more rigorous expression for the mass function is given by $\beta(M) = {\rho_\text{PBH} \over \rho_c} {1\over M} \sqrt{ {2 \over \pi} } \nu_R \exp\l(-\nu_R^2/2\r) \l| {\mathrm{d} \ln\nu_R \over \mathrm{d}\ln M} \r|$ \cite{Green:2004wb, Young:2014ana, Sureda:2020vgi}, where $\nu_R\equiv\delta_c / \sigma_R$. However, the mass function of PBHs is always calculated using Eq. \eqref{beta_def} in PBH community, and we follow this convention for a conservative estimate for PBH abundance.}
\be \label{beta}
\beta(M) = 2 \int_{\delta_c}^{\infty} P(\delta_R) \mathrm{d}\delta_R
= \text{erfc}\l[ {\delta_c \over \sqrt{2} \sigma_R} \r] ~,
\ee
where erfc is the complementary error function, and we use the Gaussian distribution function of density fluctuations $P(\delta_R) = { 1 \over \sqrt{2 \pi} \sigma_R } \exp\l( -\delta_R^2/(2 \sigma_R^2) \r)$ in Eq. \eqref{beta}. Note that $\delta_R$ is the smoothed density field, $\delta_R(\mathbf{x}) \equiv \int \mathrm{d}^3\mathbf{x}' W(\mathbf{x}- \mathbf{x}';R) \delta(\mathbf{x}')$, where $W(\mathbf{x}- \mathbf{x}';R)$ is a window function with a characteristic comoving smoothing scale $R$, and the associated mass is given by $M \equiv {4\pi\over 3} \rho_c (a R)^3$. In the case of PBH formation, the comoving smoothing scale is typically chosen as the comoving Hubble radius, namely $R = (a H)^{-1}$, and is related to the Horizon-crossing $k$ mode of density perturbations through $M \simeq M_\odot \l( k / 1.9 \times 10^6~\text{Mpc}^{-1} \r)^{-2}$ \cite{Sasaki:2018dmp}, where $M_\odot \simeq 2 \times 10^{33} \text{~g}$ is the solar mass.
In this paper, we choose the spherically symmetric real-space top-hat window function\footnote{Note that the factor $(2\pi)^{-3/2}$ appearing here is due to our Fourier transformation convention.}:
\bl
W(r;R) = {3 \over 4 \pi R^3} \Theta(R - r) ~,
\quad
W(k;R) = {1\over (2\pi)^{3/2}} {3 \l[ \sin(kR) - kR \cos(kR) \r] \over (kR)^3} ~,
\el
where $\Theta$ is the Heaviside step function.
Reference \cite{Ando:2018qdb} demonstrates that the required amplitude of density perturbations for a fixed PBH abundance is smallest for the real-space top-hat window function, as opposed to the Gaussian and k-space top-hat window functions. The variance of the smoothed density field is calculated as
\be
\sigma_R^2 \equiv \langle\delta_R^2\rangle = {1 \over (2\pi)^3}\int {\mathrm{d}k \over k} W(kR)^2 \mathcal{P}_\delta(k) ~,
\ee
where $\mathcal{P}_\delta(k) \equiv {k^3\over 2\pi^2} \int \mathrm{d}^3\mathbf{r} e^{- i \mathbf{k} \cdot \mathbf{r}} \langle \delta(\mathbf{x}) \delta(\mathbf{x} + \mathbf{r}) \rangle$, and we obtain the relationship $\mathcal{P}_\delta(k) = {16\over81} \l({k\over aH}\r)^4 \mathcal{P}_\mathcal{R}(k)$ using Eq. \eqref{delta_R}. Assuming the adiabatic background expansion after PBH formation, one can relate the initial PBH abundance to the current energy fraction \cite{Sasaki:2018dmp},
\be
f_\text{PBH}(M)
\equiv { \Omega_\text{PBH} \over \Omega_\text{DM} }
\simeq 2.7 \times 10^8 \l( \frac{M}{M_{\odot}} \r)^{-1/2} \beta(M) ~.
\ee
where $\Omega_\text{PBH}$ and $\Omega_\text{DM}$ are the normalized PBH and dark matter energy densities, respectively.
We plot $f_\text{PBH}(M)$ in Fig. \ref{fig:pbh} in terms of three parameter sets listed in Table \ref{tab:parameter}. For the parameter set 1, PBHs produced in our model can be the whole dark matter, while for the parameter set 3, PBHs can account for the LIGO/Virgo GW events. The produced PBHs in terms of the parameter set 2 can explain OGLE ultrashort-timescale microlensing events \cite{Niikura:2019kqi}.

\subsection{SIGWs and multi-frequency GW experiments}

\begin{figure}[h]
	\centering
	\includegraphics[width=0.45\textheight]{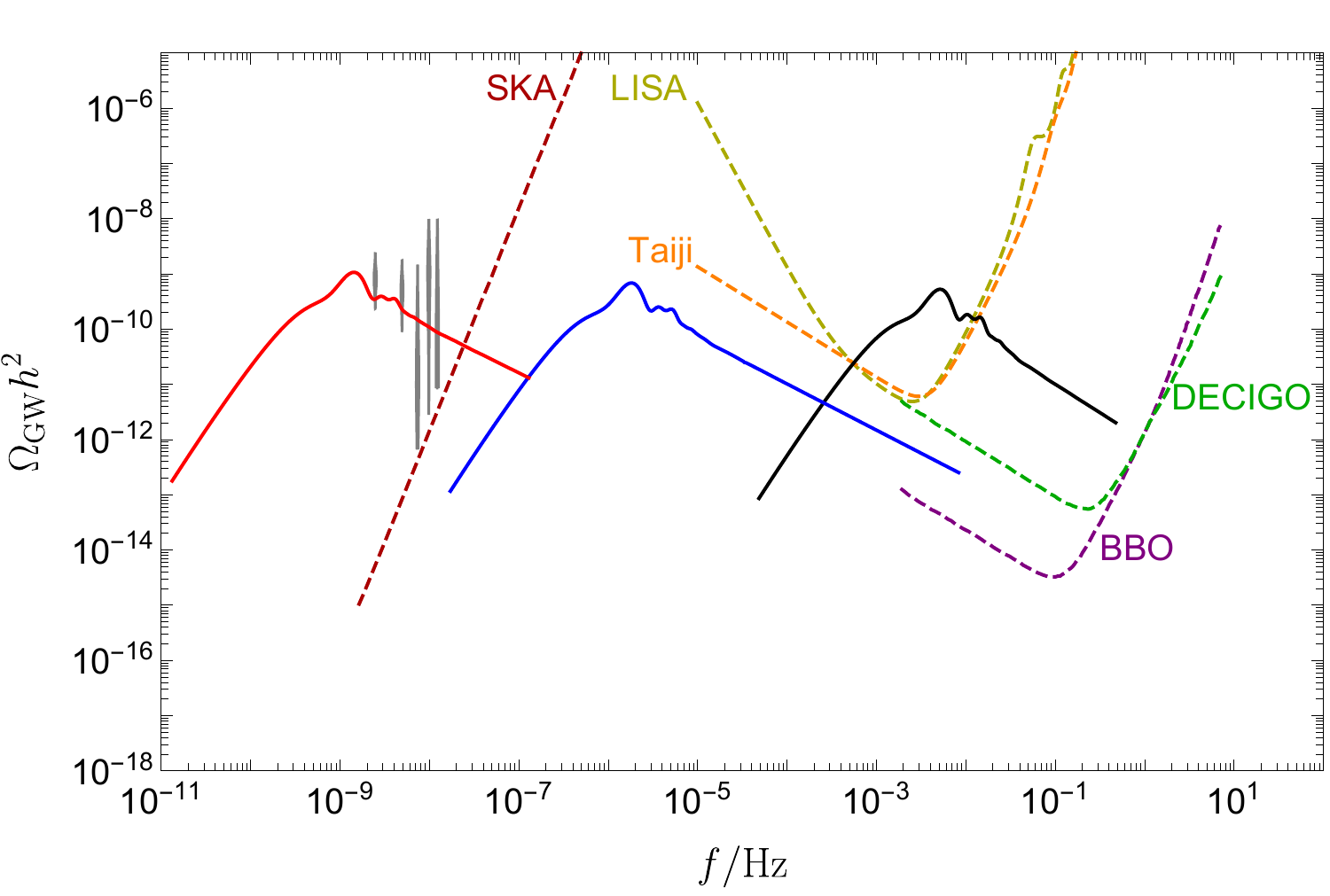}
	\caption{The current energy spectra $\Omega_{\rm GW}(\tau_0, f)$ of the SIGWs in terms of three parameter sets listed in Table \ref{tab:parameter}, with various expected sensitivity curves of the forthcoming GW experiments, e.g. SKA \cite{Janssen:2014dka}, LISA \cite{LISA:2017pwj}, Taiji \cite{Ruan:2018tsw}, DECIGO \cite{Kawamura:2011zz} and BBO \cite{Phinney:2003}. The gray vertical bars refer to the NANOGrav detection data \cite{NANOGrav:2020bcs}.}
	\label{fig:igw}	
\end{figure}

The overlarge density perturbations also induce the SIGWs when their modes re-enter the Hubble horizon during the radiation domination (RD). There is extensive literature on this direction, so that we will present the master equations and the details of their derivations can be found, e.g., in Refs.~\cite{Ananda:2006af, Baumann:2007zm, Saito:2008jc, Inomata:2016rbd, Kohri:2018awv, Espinosa:2018eve, Bartolo:2018rku, Cai:2019jah, Inomata:2020cck, Domenech:2021ztg, Cai:2021wzd, Chen:2022qec, Zhao:2022kvz, Balaji:2022dbi}. In the Newtonian gauge, the perturbed metric is written as
\be \label{metric_pert}
\mathrm{d}s^2 = a(\tau)^2 \l\{ - (1+2\Phi)\mathrm{d}\tau^2 + \l[ (1-2\Phi) \delta_{ij} + \frac{1}{2} h_{ij} \r] \mathrm{d}x^i \mathrm{d}x^j \r\} ~,
\ee
where the first-order scalar perturbation is described by $\Phi$ and we ignore the anisotropic stress at the linear order for simplicity. The secondary tensor modes $h_{ij}$, namely the SIGWs, can be induced by the nonlinear couplings between $\Phi$. The dynamics of SIGWs in Fourier space is given by
\be \label{eom_hk}
h_{\textbf{k}}^{\lambda\prime\prime}(\tau) + 2\mathcal{H}h_{\textbf{k}}^{\lambda\prime}(\tau) + k^2h_{\textbf{k}}^{\lambda}(\tau) = S^\lambda_\textbf{k}(\tau) ~,
\ee
where the source term $S^\lambda_\textbf{k}(\tau)$ during RD is given by
\be
\begin{aligned}
S^\lambda_{\mathbf{k}}(\tau)
=
4 \int {\mathrm{d}^3\mathbf{p} \over (2\pi)^{3/2}} \mathbf{e}^\lambda(\mathbf{k},\mathbf{p})
\Big[&
3 \Phi_\mathbf{p}(\tau) \Phi_{\mathbf{k} - \mathbf{p}}(\tau)
+ \mathcal{H}^{-2} \Phi_\mathbf{p}'(\tau) \Phi_{\mathbf{k} - \mathbf{p}}'(\tau)
\\&
+ \mathcal{H}^{-1} \Phi_\mathbf{p}'(\tau) \Phi_{\mathbf{k} - \mathbf{p}}(\tau)
+ \mathcal{H}^{-1} \Phi_\mathbf{p}(\tau) \Phi_{\mathbf{k} - \mathbf{p}}'(\tau)
\Big] ~,
\end{aligned}
\ee
where $\mathbf{e}^\lambda(\mathbf{k},\mathbf{p}) \equiv e^\lambda_{lm}(\mathbf{k}) p_l p_m$.
Using the Green function method, $h_{\textbf{k}}^{\lambda}(\tau) = \int^\tau \mathrm{d}\tau^\prime g_{k}(\tau,\tau^\prime)S^\lambda_\textbf{k}(\tau^\prime)$, to solve the EoM \eqref{eom_hk}, one obtains the total power spectrum (i.e., $\mathcal{P}_h = \sum_{\lambda=+,\times} \mathcal{P}_h^\lambda$) for SIGWs as
\be
\mathcal{P}_h(\tau,k) 
= \int^\infty_0 \mathrm{d}v \int^{|1+v|}_{|1-v|}\mathrm{d}u \l[ {4v^2-(1+v^2-u^2)^2 \over 4uv} \r]^2 I^2(v,u,x) \mathcal{P}_\mathcal{R}(ku) \mathcal{P}_\mathcal{R}(kv) ~.
\ee
where we define $u \equiv |\mathbf{k} - \mathbf{p}|/k$, $v \equiv p/k$ and $x \equiv k\tau$. The kernel function $I^2(v,u,x)$ describes the evolution information of the source and can be calculated as \cite{Kohri:2018awv, Espinosa:2018eve},
\be
\begin{aligned}
\overline{I^2(v,u,x\rightarrow \infty)} =& 
4 \left( \frac{3(u^2+v^2-3)}{4 u^3 v^3 x}\right)^2 \bigg[ \bigg(-4uv+(u^2+v^2-3) \ln\left| \frac{3-(u+v)^2}{3-(u-v)^2}\right| \bigg)^2  
\\& 
+ \pi^2(u^2+v^2-3)^2 \Theta(v+u-\sqrt{3})\bigg] ~,
\end{aligned}
\ee
where the overline denotes the time average. The large limit $x\rightarrow \infty$ is taken in the above equation, as we are primarily interested in the present GW spectrum. Using the canonical definition of GW's energy density \cite{Brill:1964zz, Isaacson:1968hbi, Isaacson:1968zza, Ford:1977dj, Maggiore:1999vm, Boyle:2005se, Ota:2021fdv}, given by $\rho_{\text{GW}}(\tau,\mathbf{x}) = { M_{\text{pl}}^2 \over 16 a^2(\tau) } \langle h'_{ij}(\tau,\mathbf{x}) h^{ij}{}'(\tau,\mathbf{x}) \rangle$\footnote{The prefactor $1/2$ in the metric perturbations \eqref{metric_pert} is counted in the total GW energy.}, the relationship of the GW energy spectrum and its power spectrum is written as,
\be \label{omega_gw}
\Omega_{\rm GW}(\tau,k) 
= \frac{1}{48} \l( {k \over a H} \r)^2 \overline{\mathcal{P}_h(\tau,k)} ~.
\ee
The GW energy density starts to decay relative to matter energy density after the radiation-matter equality $\tau_\text{eq}$, the GW spectrum observed today is thus given by \cite{Pi:2020otn}
\be
\Omega_{\rm GW}(\tau_0, f) h^2 \simeq 1.6 \times 10^{-5} \l( {g_{\ast,s} \over 106.75} \r)^{-1/3} \l( { \Omega_{\rm r,0} h^2 \over 4.1 \times 10^{-5} } \r) \Omega_{\rm GW}(\tau_\text{eq}, f) ~,
\ee
where $\Omega_{\rm GW}(\tau_\text{eq}, f)$ can be calculated by Eq. \eqref{omega_gw} at the radiation-matter equality, with the physical frequency $f = k/(2\pi a_0) \simeq 1.5 \times 10^{-9} (k/1 \text{~pc}^{-1}) ~\text{Hz}$. Since the scalar perturbations damp quickly on the subhorizon during radiation, a majority of SIGWs is thus produced just after the source reenters the horizon \cite{Espinosa:2018eve}.  $\Omega_{\rm GW}(\tau_0, f)$ are shown in Fig. \ref{fig:igw} in terms of three parameter sets listed in Table \ref{tab:parameter}. The predicted SIGWs are detectable by the future GW observations. As for the parameter set 3, the consequent SIGWs (red) can be matching with the NANOGrav detection data \cite{NANOGrav:2020bcs}.
Note that it is challenging to differentiate between PBH formation models based solely on PBH mass spectrum (shown in Fig. \ref{fig:pbh}), because they predict mass spectra with comparable bump-like shapes based on the Press-Schechter formalism \eqref{beta}. Nonetheless, it is possible to break this degeneracy by examining the concomitant SIGWs, which depend on the shape of the curvature power spectrum and different PBH formation models typically predict distinct curvature spectra (refer to Refs. \cite{Domenech:2021ztg, Escriva:2022duf} for detailed discussions).

\section{Conclusions}
\label{sec:conclusion}

PBHs are a promising tool for probing early Universe physics, particularly for detecting small-scale nontrivial physical phenomena. The amplification of PBH formation can be realized within the framework of the inflationary Universe by introducing extra fields or non-trivial background dynamics. One direct way is to slow down the inflaton's speed in a short period, which is the case for a flatten or a bump-like potential. In this paper, we develop a two-field inflation with a smooth Gaussian potential which acts as a two-dimensional bump on the top of the $\phi^{2/5}$ potential, the braking and turning of the inflationary trajectory can be easily realized when the inflaton encounters this potential bump. 
Compared to previous works on similar topics, one advantage of using the two-dimensional Gaussian potential is that it requires fewer fine-tunings of initial conditions. This allows the inflaton to generally undergo braking phases, which can result in the growth of comoving curvature perturbations.
In our analysis, we perform matching calculations to the braking phase (approximated by five successive CR phases) and obtain estimates of several features of the curvature power spectrum at the end of inflation, including the growth rate, dip scale, and peak amplitude. Our results indicate that the braking phase provides the primary contribution to the spectrum, while the turning phase contributes additional spikes around the peak amplitude. However, as the duration of the turning phase is too short, its impact on the overall spectrum is not significant.

Our paper provides a simple example of generalizing a single-field bump-like potential to a two-field inflation model, which can alleviate the fine-tuning issue to some extent. There are several extensions of this work that require further investigation. For example, recent studies have shown that quantum \cite{Ota:2022hvh,Inomata:2022yte, Kristiano:2022maq} or classical \cite{Chen:2022dah} one-loop corrections can significantly influence PBH formation and SIGWs. Similar one-loop corrections can also be considered in our model. 
Additionally, we have ignored non-Gaussianity and quantum stochastic effects in our analysis for simplicity. Future studies could include these effects to obtain more precise results for PBH formation with our Gaussian bump potential.
Moreover, if the Gaussian bump is symmetric or slightly asymmetric with respect to the inflationary trajectory, both trajectories $A$ and $B$ shown in Fig. \ref{fig:ms} need to be considered, which may also affect the PBH abundance and their spatial clustering. We leave these interesting questions in the follow-up works.

%%%%%%%%%%%%%%%%%%%%%%%%%%%%%%%%%%%%%%%%%%%%%%%%%%%%%%%%%%%%%%%%%%%%%%%%%%%%%%%%
\begin{acknowledgments}

We are grateful to Xiao-Han Ma and Yi Wang for valuable discussions.
We thank the referee for valuable suggestions.
C.C. is grateful to Yudong Luo for encouraging discussions on optimization algorithms.
C.C. thanks the Particle Cosmology Group at University of Science and Technology of China during his visit. This work is supported in part by the National Key R\&D Program of China (No. 2021YFC2203100).
C.C. is supported by the Jockey Club Institute for Advanced Study at The Hong Kong University of Science and Technology.

\end{acknowledgments}

\appendix

\section{The matching calculations for constant-$\eta$ phases}
\label{app:match}

In what follows, we present analytical matching calculations for sudden transitions between various constant-$\eta$ phases, mainly follow the treatments in Ref. \cite{Byrnes:2018txb}, and similar calculations can be found in Refs. \cite{Carrilho:2019oqg, Tasinato:2020vdk}. The Mukhanov-Sasaki equation for comoving curvature perturbations is written as
$v_k'' + \l( k^2 - z''/z \r) v_k = 0$, where $v_k = z \mathcal{R}_k$ and $z^2 = 2 a^2 M_\text{Pl}^2 \epsilon$. It can be shown that
\be
{z'' \over z} = (a H)^2 \l( 2 - \epsilon + {3\over2} \eta + {1\over4} \eta^2 - {1\over2} \epsilon\eta + {1\over2} {\dot{\eta}\over H} \r) ~,
\ee
which is the exact expression to all orders. Considering $\epsilon \ll 1$ and $\eta$ a constant, we drop higher-order terms, the Mukhanov-Sasaki equation becomes
\be
v_k'' + \l( k^2 - {\nu^2 - 1/4 \over \tau^2} \r) v_k = 0 ~,
\ee
where $\nu = {3 + \eta\over2}$. Then, the general solution of $\mathcal{R}_k(\tau)$ is given by
\be \label{sol_rk}
\mathcal{R}_k(\tau) =
{ \sqrt{-\tau}  \over a(\tau) M_\text{Pl} \sqrt{2 \epsilon} } \l[ A H_\nu^{(1)}(- k \tau) + B H_\nu^{(2)}(- k \tau) \r] ~,
\ee
where the constants $A$ and $B$ are determined by initial conditions. For example, the initial condition is chosen be the Bunch-Davies vacuum state, namely $\lim_{\tau\rightarrow-\infty}  v_k \simeq {1 \over \sqrt{2 k}} e^{- i k \tau}$, and then we obtain
\be \label{rk_bd}
\mathcal{R}_k(\tau) = {\sqrt{\pi}\over2 a(\tau) M_\text{Pl} \sqrt{2 \epsilon} } e^{i \l( \nu + {1\over2} \r) {\pi\over2} } \sqrt{-\tau} H_\nu^{(1)}(- k \tau) ~,
\ee
where $a(\tau)=-1/(H\tau)$. Within each constant-$\eta$ phase, the corresponding coefficients $A$ and $B$ in the solution \eqref{sol_rk} are determined by the matching condition with the previous phase. We expect that there is no energy jump in the background between two phases, so the comoving curvature perturbation and its first time derivative should be continuous at the transition point \cite{Byrnes:2018txb},
\be \label{matching}
[\mathcal{R}_k]_{\pm} = 0 ~,
\quad
[\mathcal{R}_k']_{\pm} = 0 ~,
\ee 
which is so called the Israel junction condition \cite{Israel:1966rt,Deruelle:1995kd}. The final analytical results of curvature power spectrum for the transitions $\eta = 0 \rightarrow \eta = -3 \rightarrow \eta = 0.5 \rightarrow \eta = 2 \rightarrow \eta = 0$ are quite lengthy and not particularly meaningful to present in exact expressions. Without loss of generality, we will present the formalism for two typical sudden transitions $0 \rightarrow \eta$ and $\eta_1 \rightarrow \eta_2$.

{\it Transition $0 \rightarrow \eta$:} In the first SR phase, the evolution of comoving curvature perturbation derived from Eq. \eqref{rk_bd} is given by
\be \label{sr_sol}
\mathcal{R}_k(\tau) 
=  i {H \over M_\text{Pl}} { e^{-ik\tau}\over \sqrt{4 \epsilon_1 k^3 } } (1+ik\tau) ~,
\ee
where $\epsilon_1$ is the SR parameter during this SR phase. The solution of $\mathcal{R}_k$ in the second $\eta$-phase is given by Eq. \eqref{sol_rk} with the corresponding SR parameter $\epsilon_2(\tau) = \epsilon_1 \l( \tau_1/\tau \r)^\eta$, where $\tau_1$ is the time of the sudden transition. Using the matching condition \eqref{matching}, we calculate the coefficients as
\bl \label{A}
A &= - { e^{-i k \tau_1} \pi \over 4 \sqrt{2} \sqrt{- \tau_1} k} \l[ (1+ik\tau_1) H_{{1+\eta \over 2}}^{(2)}(- k \tau_1) + k \tau_1 H_{{3+\eta \over 2}}^{(2)}(- k \tau_1) \r] ~,
\\ \label{B}
B &= { e^{-i k \tau_1} \pi \over 4 \sqrt{2} \sqrt{- \tau_1} k} \l[ (1+ik\tau_1) H_{{1+\eta \over 2}}^{(1)}(- k \tau_1) + k \tau_1 H_{{3+\eta \over 2}}^{(1)}(- k \tau_1) \r] ~.
\el

{\it Transition $\eta_1 \rightarrow \eta_2$:} In both of these phases, the solutions are given by Eq. \eqref{sol_rk} and we use subscripts to distinguish coefficients in each phase, i.e., $(A_1, B_1)$ and $(A_2, B_2)$. Using the junction condition \eqref{matching}, we obtain
\be {\scriptsize
\begin{aligned}
A_2 &= {1\over k \tau_1} \left[\left(H_{\frac{\eta_2+1}{2}}^{(1)}(-k
\tau_1)-H_{\frac{\eta_2+5}{2}}^{(1)}(-k \tau_1)\right) H_{\frac{\eta_2+3}{2}}^{(2)}(-k \tau_1)+H_{\frac{\eta_2+3}{2}}^{(1)}(-k \tau_1)
\left(H_{\frac{\eta_2+5}{2}}^{(2)}(-k \tau_1)-H_{\frac{\eta_2+1}{2}}^{(2)}(-k \tau_1)\right)\right]^{-1}
\\& \times
\begin{bmatrix}
	&A_1 H_{\frac{\eta_1+3}{2}}^{(1)}(-k \tau_1) \left(-k
	\tau_1 H_{\frac{\eta_2+1}{2}}^{(2)}(-k \tau_1)+\eta_2 H_{\frac{\eta_2+3}{2}}^{(2)}(-k \tau_1)+k \tau_1 H_{\frac{\eta_2+5}{2}}^{(2)}(-k \tau_1)\right)
	\\&
	+k \tau_1 H_{\frac{\eta_2+3}{2}}^{(2)}(-k \tau_1) \left(A_1 H_{\frac{\eta_1+1}{2}}^{(1)}(-k
	\tau_1)-A_1 H_{\frac{\eta_1+5}{2}}^{(1)}(-k \tau_1)+B_1 H_{\frac{\eta_1+1}{2}}^{(2)}(-k \tau_1)-B_1 H_{\frac{\eta_1+5}{2}}^{(2)}(-k \tau_1)\right)
	\\&
	+B_1 H_{\frac{\eta_1+3}{2}}^{(2)}(-k
	\tau_1) \left(-k \tau_1 H_{\frac{\eta_2+1}{2}}^{(2)}(-k \tau_1)+\eta_2 H_{\frac{\eta_2+3}{2}}^{(2)}(-k \tau_1)+k \tau_1 H_{\frac{\eta_2+5}{2}}^{(2)}(-k \tau_1)\right)
\end{bmatrix} ~.
\end{aligned} }
\ee

\be {\scriptsize
\begin{aligned}
B_2 &= {1\over k \tau_1} \left[\left(H_{\frac{\eta_2+1}{2}}^{(1)}(-k
\tau_1)-H_{\frac{\eta_2+5}{2}}^{(1)}(-k \tau_1)\right) H_{\frac{\eta_2+3}{2}}^{(2)}(-k \tau_1)+H_{\frac{\eta_2+3}{2}}^{(1)}(-k \tau_1)
\left(H_{\frac{\eta_2+5}{2}}^{(2)}(-k \tau_1)-H_{\frac{\eta_2+1}{2}}^{(2)}(-k \tau_1)\right)\right]^{-1}
\\& \times
\begin{bmatrix}
	&-A_1 k \tau_1 H_{\frac{\eta_1+1}{2}}^{(1)}(-k \tau_1) H_{\frac{\eta_2+3}{2}}^{(1)}(-k \tau_1)+A_1 k
	\tau_1 H_{\frac{\eta_1+5}{2}}^{(1)}(-k \tau_1)
	H_{\frac{\eta_2+3}{2}}^{(1)}(-k \tau_1)
	\\&
	+A_1
	H_{\frac{\eta_1+3}{2}}^{(1)}(-k \tau_1) \left(k \tau_1 H_{\frac{\eta_2+1}{2}}^{(1)}(-k \tau_1)-\eta_2
	H_{\frac{\eta_2+3}{2}}^{(1)}(-k \tau_1)-k \tau_1
	H_{\frac{\eta_2+5}{2}}^{(1)}(-k \tau_1)\right)
	\\&
	-k B_1 \tau_1 H_{\frac{\eta_1+1}{2}}^{(2)}(-k \tau_1)
	H_{\frac{\eta_2+3}{2}}^{(1)}(-k \tau_1) 
	-\eta_2	B_1 H_{\frac{\eta_1+3}{2}}^{(2)}(-k \tau_1)
	H_{\frac{\eta_2+3}{2}}^{(1)}(-k \tau_1)
	+k B_1 	\tau_1 H_{\frac{\eta_1+5}{2}}^{(2)}(-k \tau_1)
	\\&
	H_{\frac{\eta_2+3}{2}}^{(1)}(-k \tau_1)
	+k B_1 \tau_1 H_{\frac{\eta_1+3}{2}}^{(2)}(-k \tau_1)
	H_{\frac{\eta_2+1}{2}}^{(1)}(-k \tau_1)
	-k B_1 \tau_1 H_{\frac{\eta_1+3}{2}}^{(2)}(-k \tau_1)
	H_{\frac{\eta_2+5}{2}}^{(1)}(-k \tau_1)
\end{bmatrix} ~.	
\end{aligned} }
\ee
With the choice of parameters $(A_1, B_1) = (- {\sqrt{\pi}\over2}, 0 )$ [i.e., referring to the SR solution \eqref{sr_sol}], the above two coefficients indeed reduce to the expressions \eqref{A} and \eqref{B}.
As mentioned earlier, inflation needs to return to the SR phase prior to its end. For this last SR phase, it is conventional to rewrite the solution \eqref{sol_rk} as
\be
\mathcal{R}_k(\tau) 
= i {H \over M_\text{Pl}} { 1 \over \sqrt{4 \epsilon k^3 } } \l[ A (1+ik\tau) e^{-ik\tau} - B (1 - ik\tau) e^{ik\tau} \r] ~,
\ee
where $A$ and $B$ are determined through the above matching calculations, $\epsilon$ is the corresponding SR parameter. Hence, the final power spectrum can be expressed as
\be
\mathcal{P}_{\mathcal{R}}(k) 
= \lim_{\tau \rightarrow 0^{-}} {k^3 \over 2 \pi} |\mathcal{R}_k|^2
= {H^2 \over 8 \pi^2 M_\text{Pl}^2 \epsilon} |A-B|^2 ~. 
\ee

\section{A non-Gaussian potential bump} \label{app:ng}

\begin{figure}[ht!]
	\centering
	\includegraphics[width=0.28\textheight]{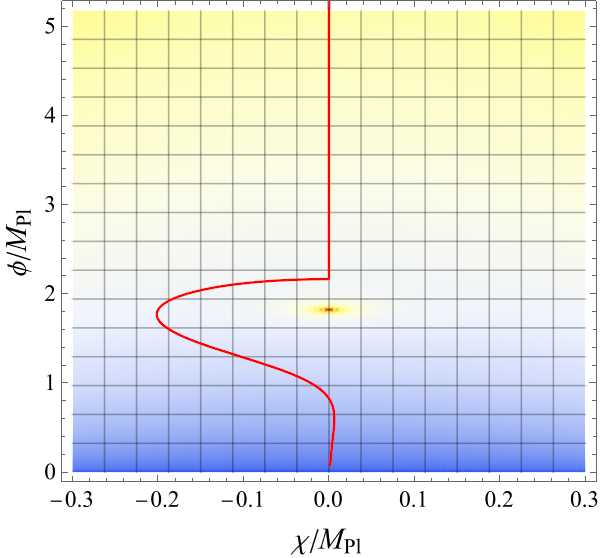}	
	\includegraphics[width=0.28\textheight]{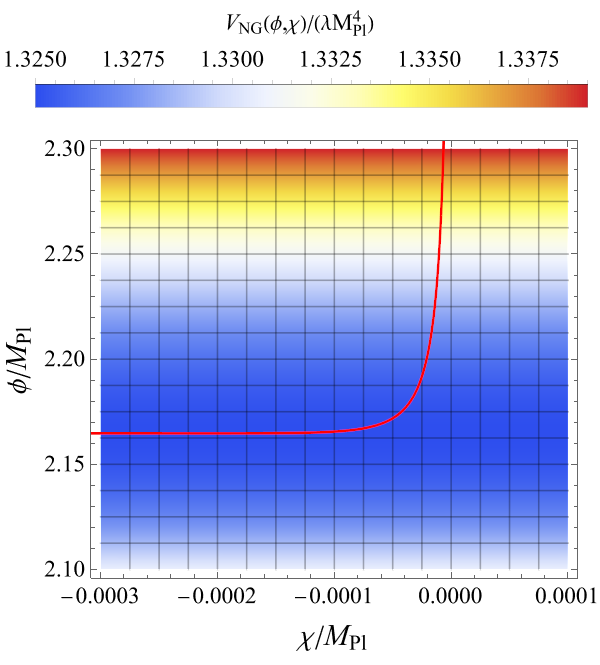}	
	\includegraphics[width=0.28\textheight]{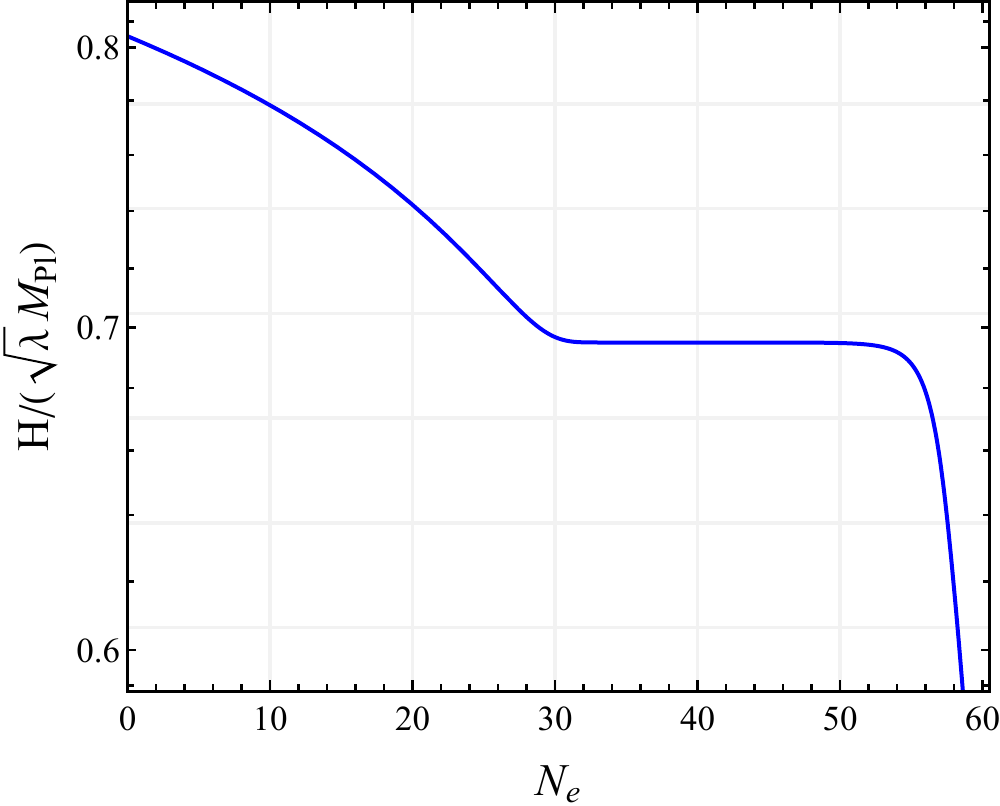}
	\includegraphics[width=0.28\textheight]{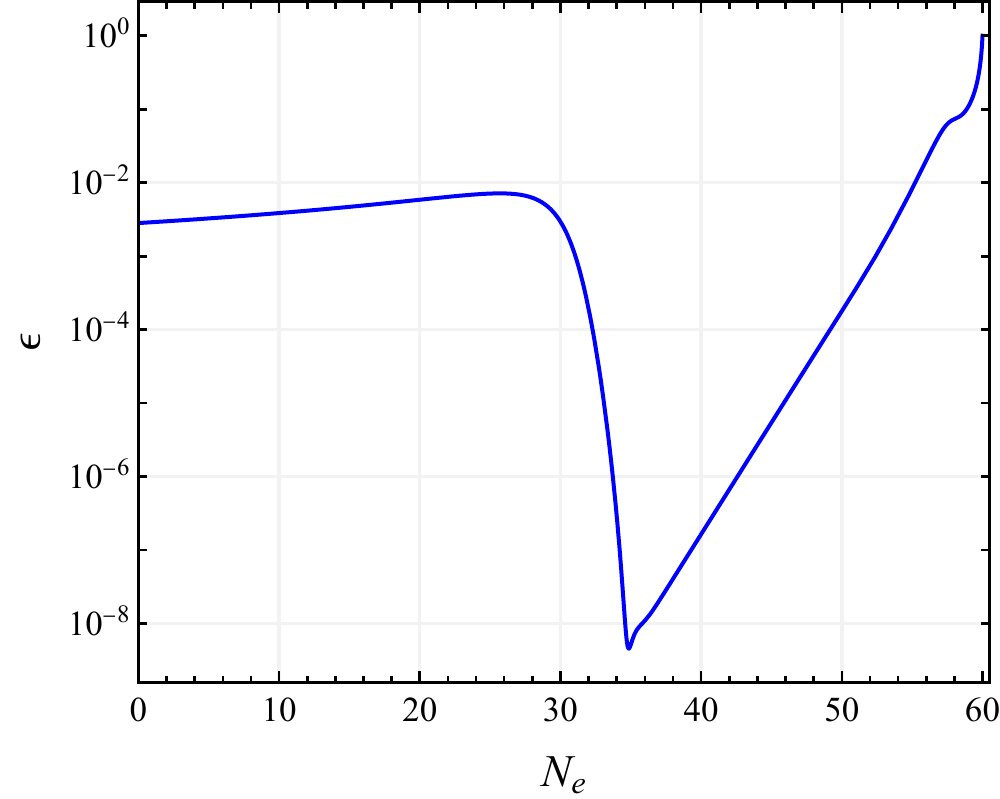}
	\includegraphics[width=0.28\textheight]{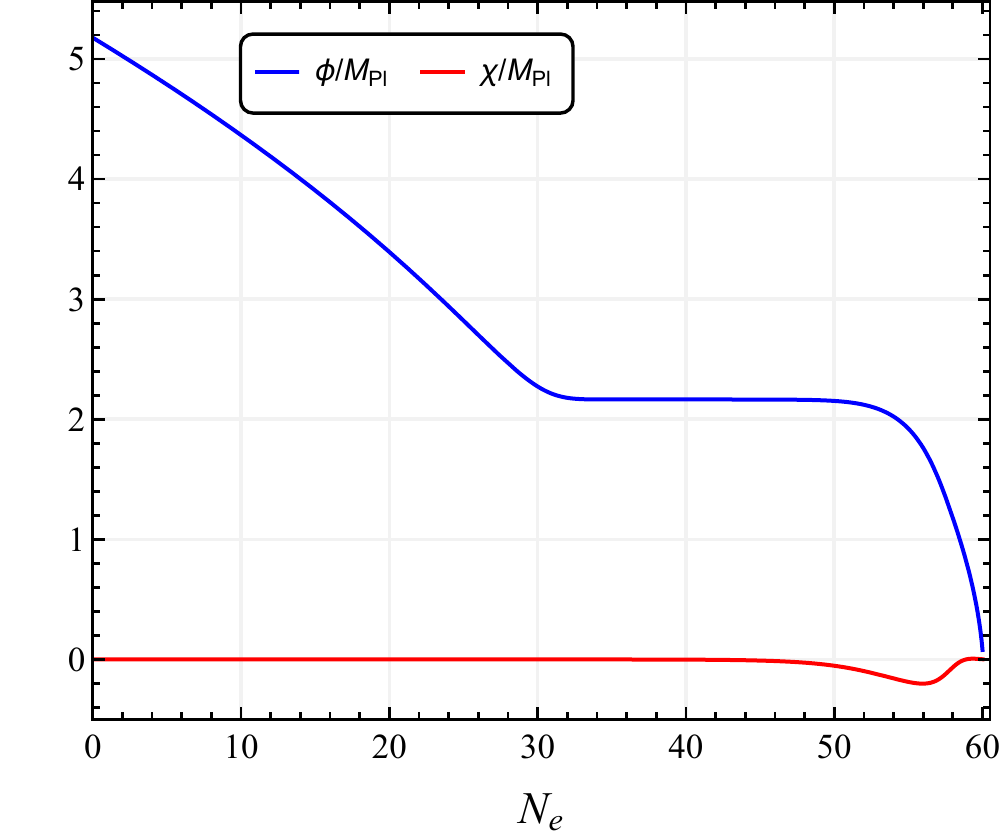}
	\includegraphics[width=0.28\textheight]{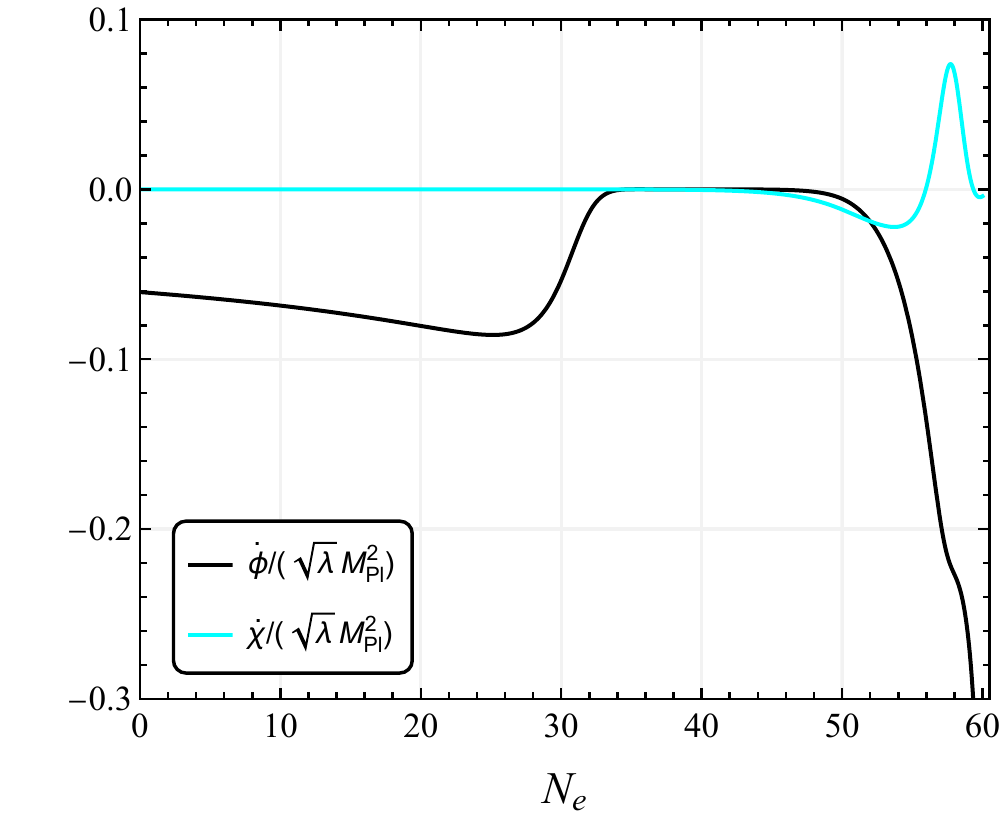}
	\caption{The background evolutions for the non-Gaussian potential bump \eqref{app:potenNG}, which corresponds to Fig. \ref{fig:bg} for the Gaussian potential bump \eqref{eq:potential}. The parameters are taken as: $m^2/(\lambda M_{\text{Pl}}^2) =2$, $\Lambda^4/(\lambda M_{\text{Pl}}^4) =2$, $\phi_c/M_{\text{Pl}}=1.82$, $\chi_c/M_{\text{Pl}} = 1.588 \times 10^{-5}$, $\sigma_\phi/M_{\text{Pl}} = 0.015$, $\sigma_\chi/M_{\text{Pl}} = 0.008$ and $\lambda = 7.11\times10^{-10}$.}
	\label{fig:bgNG}	
\end{figure}

\begin{figure}[ht!]
	\centering
	\includegraphics[width=0.45\textheight]{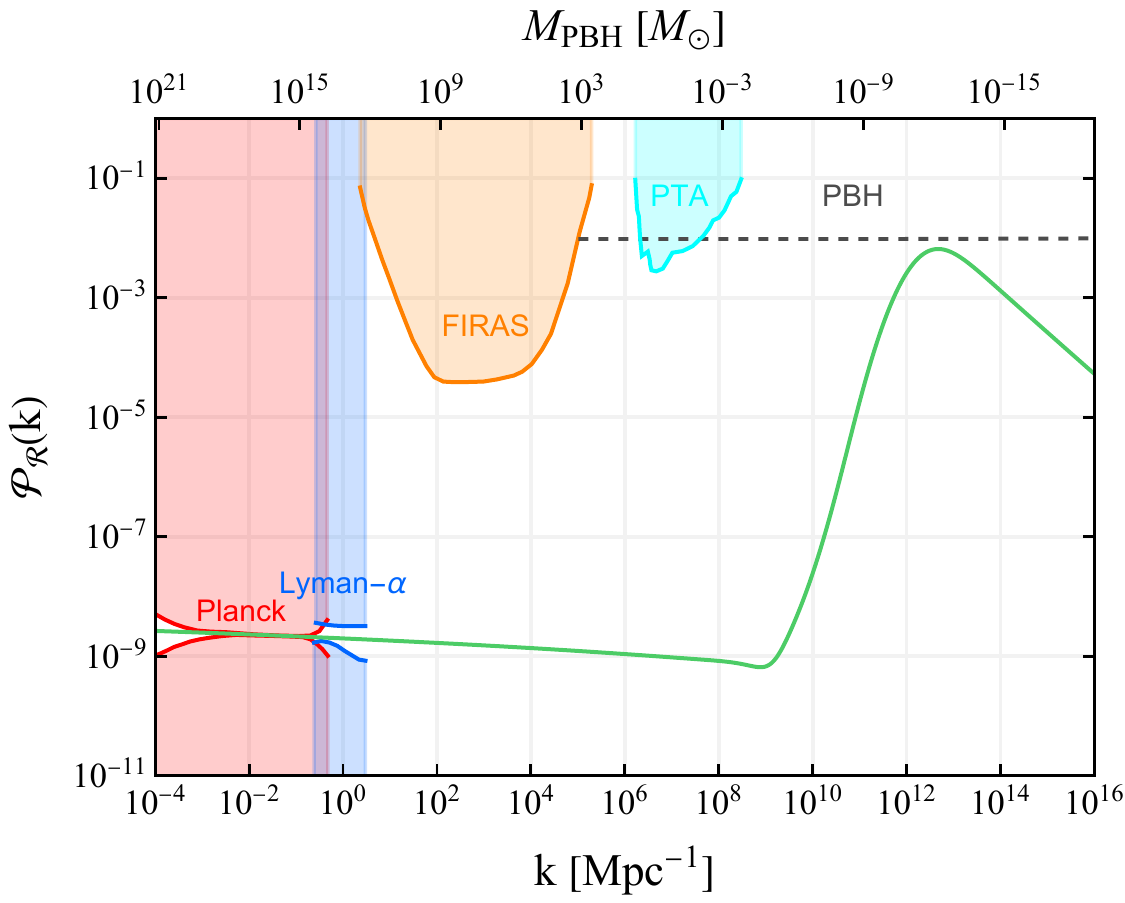}
	\caption{The numerical result of the curvature power spectrum at the end of inflation for the non-Gaussian potential bump \eqref{app:potenNG}, which corresponds to Fig. \ref{fig:sp} for the Gaussian potential bump \eqref{eq:potential}.}
	\label{fig:spNG}
\end{figure}

\begin{figure}[ht!]
	\centering
	\includegraphics[width=0.45\textheight]{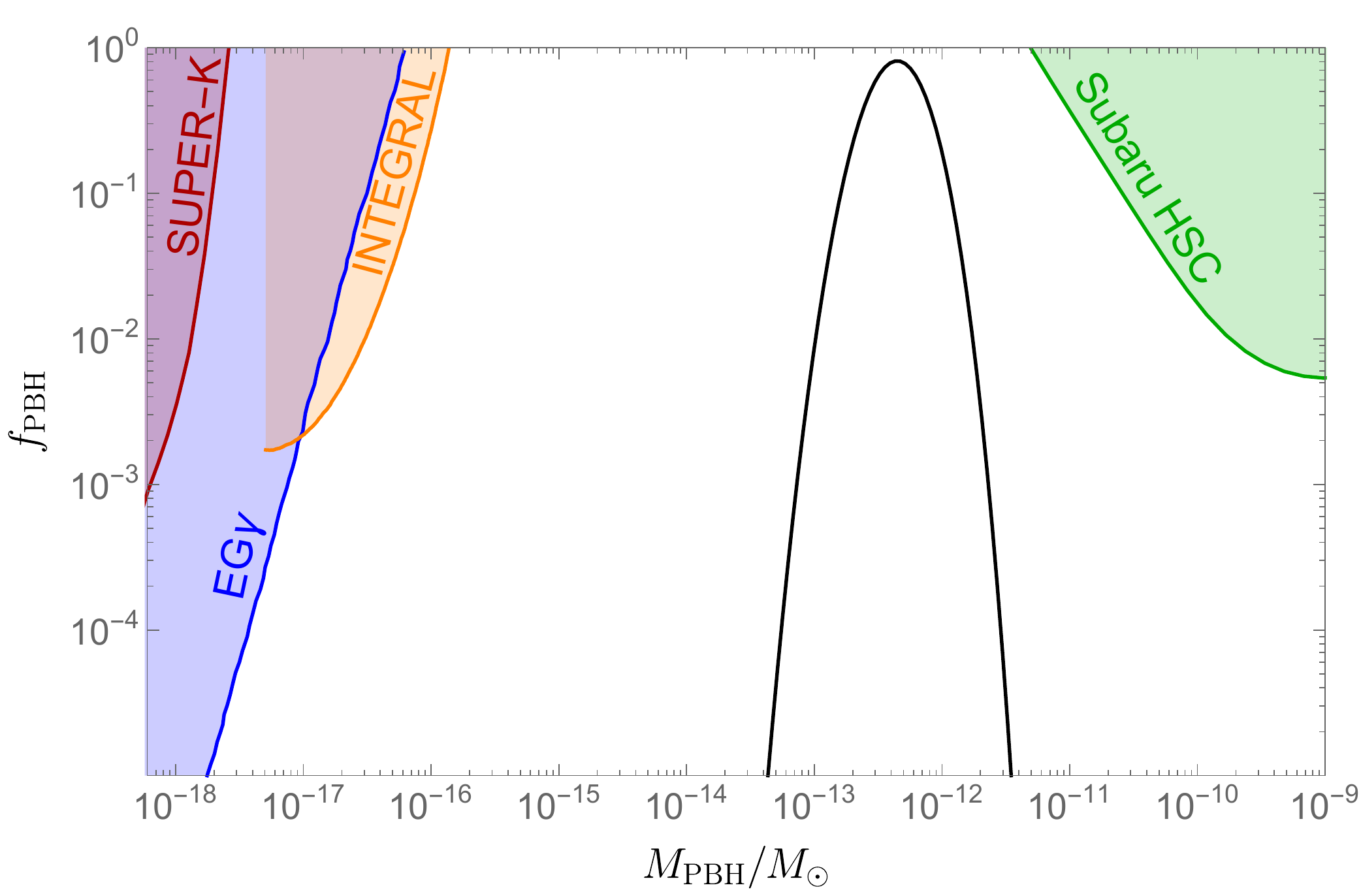}
	\caption{The current energy fraction of PBHs $f_\text{PBH}(M)$ for the non-Gaussian potential bump \eqref{app:potenNG}, which corresponds to Fig. \ref{fig:pbh} for the Gaussian potential bump \eqref{eq:potential}.}
	\label{fig:pbhNG}	
\end{figure}

\begin{figure}[hbt!]
	\centering
	\includegraphics[width=0.45\textheight]{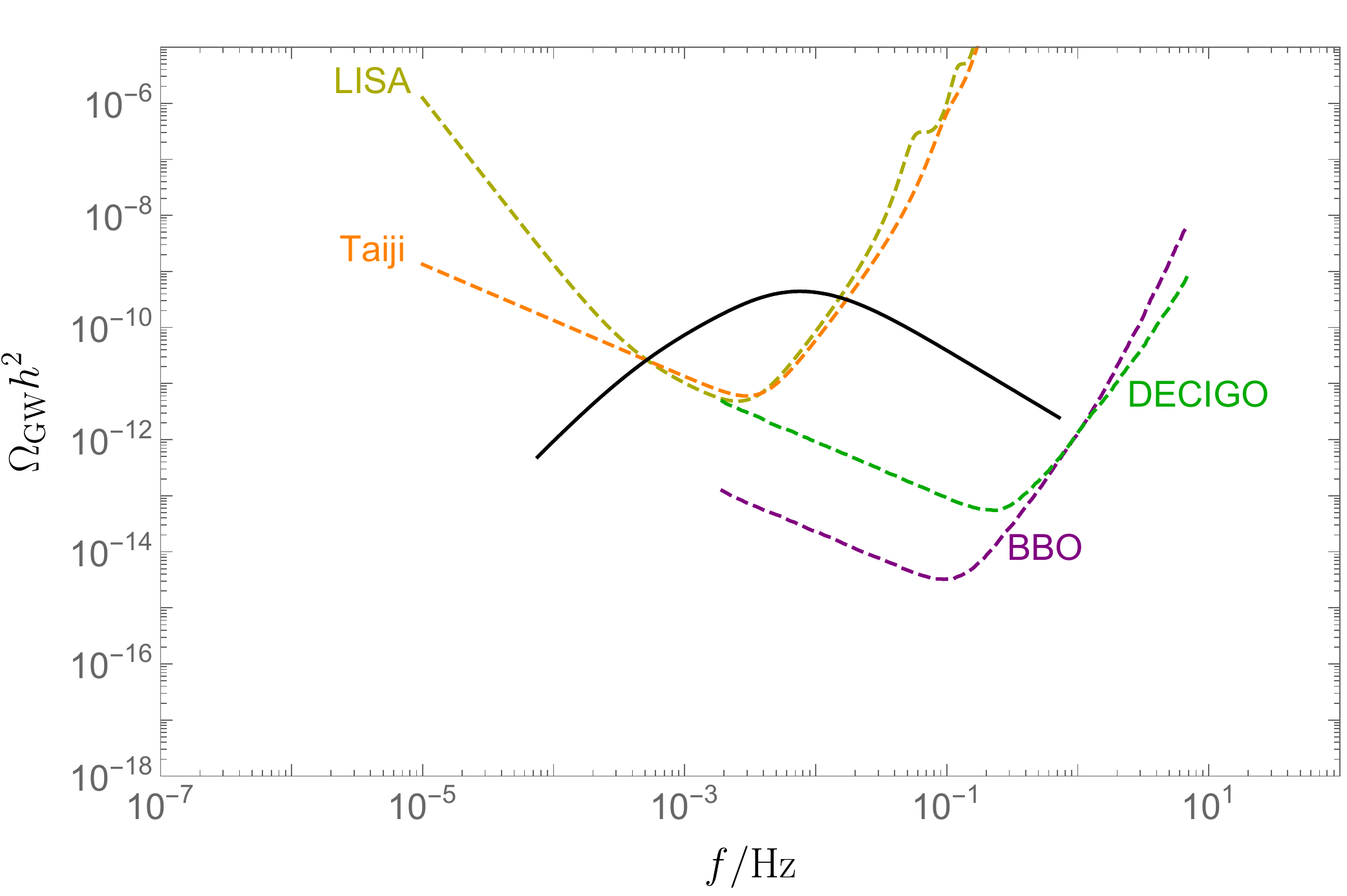}
	\caption{The current energy spectra $\Omega_{\rm GW}(\tau_0, f)$ of the SIGWs for the non-Gaussian potential bump \eqref{app:potenNG}, which corresponds to Fig. \ref{fig:igw} for the Gaussian potential bump \eqref{eq:potential}.}
	\label{fig:igwNG}
\end{figure}

To further illustrate the possibility of enhancing curvature perturbation with a two-dimensional potential bump, we present a concrete non-Gaussian bump on the top of the same basis potential $V_\text{basis}(\phi, \chi)$ as Eq. \eqref{eq:potential}. The total potential is written as
\be \label{app:potenNG}
V_\text{NG}(\phi, \chi) = V_\text{basis}(\phi, \chi) + {\Lambda^4 \over \l[ 1 + {(\phi - \phi_c)^2 \over \sigma_\phi^2} + {(\chi - \chi_c)^2 \over \sigma_\chi^2} \r]^{1/2} } ~.
\ee
Performing numerical calculations based on the same background equations \eqref{eq:bg_phi}, \eqref{eq:bg_chi}, and \eqref{eq:hubble} as those used for Fig. \ref{fig:bg}, we obtained the background evolutions shown in Fig. \ref{fig:bgNG}. It is evident from Fig. \ref{fig:bgNG} that the inflaton's speed is reduced as it encounters the non-Gaussian bump similar to the case of the Gaussian bump discussed previously. The amplification of curvature perturbation is also observed 
as illustrated in Fig. \ref{fig:spNG}, which also satisfies the CMB constraint. The scalar power index and the tensor-to-scalar ratio are given by $n_s = 0.9675$ and $0.046$, respectively. 
The resulting power spectrum of curvature perturbations, the PBH current abundance and the SIGW spectrum are shown in Figs. \ref{fig:spNG}, \ref{fig:pbhNG} and \ref{fig:igwNG}, respectively.
It is anticipated that the realization of PBH formation can be achieved by devising a suitable basis potential $V_\text{basis}$ along with a potential bump $V_\text{bump}$.

%%%%%%%%%%%%%%%%%%%%%%%%%%%
\bibliographystyle{plain}
\bibliography{PBHTwoField}
%%%%%%%%%%%%%%%%%%%%%%%%%%%

\end{document}